\documentstyle[12pt,aaspp4]{article}
\begin{document}
\title{The published extended rotation curves of spiral galaxies:
Confrontation with Modified Dynamics} 
\author{R.H. Sanders}
\affil{Kapteyn Astronomical Institute, Groningen, The Netherlands}
\authoremail{sanders@astro.rug.nl}

\begin{abstract}
A sample of 22 spiral galaxy rotation curves, measured in the 21 cm line
of neutral hydrogen, is considered in the context of Milgrom's 
modified dynamics (MOND).  Combined with the previous highly selected
sample of Begeman et al. (1990), this comprises the current total sample
of galaxies with published (or available) extended rotation curves
and photometric observations of the light distribution.  
This is the observational basis of present quantitative understanding 
of the discrepancy between the visible mass and classical dynamical mass
in galaxies.  It is found that the gravitational force calculated
from the observed distribution of luminous
material and gas using the simple MOND formula
can account for the overall shape and amplitude of these 22 rotation
curves, and in some cases, the predicted curve agrees with the observed
rotation curve in detail.  The fitted rotation curves have, in 13
cases, only one free parameter which is the mass-to-light ratio of
the luminous disk;  in nine cases, there is an additional free parameter
which is M/L of a central bulge or light concentration.  The values
of the global M/L (bulge plus disk) are reasonable and, when the gas
mass is also included, show a scatter which is consistent with that
in the Tully-Fisher relation.  The success of the MOND prescription
in predicting the rotation curves in this larger, less stringently selected
sample, lends further support to the idea that dynamics or gravity
is non-Newtonian in the limit of low accelerations and that it is 
unnecessary to invoke the presence of large quantities of unseen matter.

\end{abstract} 

\section{Introduction}

The conventional explanation for the discrepancy between the Newtonian
dynamical mass and the luminous mass in galaxies is that the visible
galaxy is embedded in a more extensive dark halo.  But there are several
unconventional explanations involving modifications of the law of gravity
or inertia (see Sanders, 1990, for a review of the early suggestions).
On a phenomenological level, the most successful of these suggestions is
that of modified Newtonian dynamics (MOND) by Milgrom (1983).  Here,
the central idea is that the law of gravity 
or inertia assumes a 
specific non-standard form below a fixed, universal value of the 
acceleration, $a_o$, the one parameter of the theory (\cite{ml83}).

The rotation curves of spiral galaxies as measured in the 21 cm
line of neutral hydrogen, comprise the ideal body of data to confront 
such ideas.  This is because the rotation curves usually
extend well beyond the optical image of the galaxy where the 
discrepancy is large, and because gas on very nearly circular orbits 
is the most precise probe
of the radial force law in the limit of low acceleration.   
Rotation curves are not useful as tests of the dark matter
hypothesis because models consisting
of a luminous disk plus an extended dark halo generally have 
at least three adjustable parameters and can always be
concocted to fit the rotation curves (\cite{mhr96});  by fitting
to rotation curves one simply determines
the parameters of the assumed halo model.  Although the plausibity
of the fitted values of parameters is sometimes questionable, 
it is not possible
to definitively falsify the dark matter hypothesis in this way.  On the
other hand, MOND, with only one universal parameter which 
cannot vary from galaxy to galaxy, is far less flexible and far more
falsifiable.

The essential problem with the use of such data
for this purpose is that the measured
rotation curves are not all equally good approximations to the run
of circular velocity.  In many galaxies there are complications arising
from warping of the gas layer in the outer regions; this gives an
intrinsic uncertainty to the inclination and position angle of the
plane of the disk.  In other galaxies the observation of
highly asymmetric gas disks
calls into question the assumption of relaxed motion on circular orbits.
Apart from the rotation curves, 
an additional problem is that, by whatever theory of gravity one applies, 
the radial
force due to the detectable matter
is calculated by assuming that the distribution of visible light
is the precise tracer of matter in the stellar disk, and that the
distribution of neutral hydrogen is a tracer of the gaseous mass 
distribution.  These assumptions can fail in several respects:  For example,
when there are radial color gradients in a galaxy, not all color bands
can be equally good tracers of the visible mass distribution.
With respect to the gaseous component, the radial distribution and even the 
normalization of the mass of molecular gas is unknown in most spiral
galaxies.

With a view toward minimizing these problems, Begeman et al. (1991, hereafter
BBS), applied strict selection criteria to the rotation curves available 
at that time (about 20).  Most of these criteria are relevant to the
21 cm line observations:  the rotation curve had to be 
derived from two-dimensional high spatial resolution data
which eliminates distant galaxies (systemic velocity greater than 2000 km/s);
galaxies with highly warped or asymmetric gas disks or those with a patchy HI 
distribution were not considered.  High precision photometric
data (CCD in general) had to be available for estimating the contribution
of the visible disk to the radial force law.  

A total sample of 11 galaxies met these criteria.  
The conclusion of the rotation curve fitting was
that MOND, with a fixed value for the acceleration parameter and
with one free parameter per galaxy (M/L for the visible disk) worked
as well as multi-parameter dark halo models; in fact, with respect to fitting
details of the rotation curves, MOND worked even better in several cases.
The success of MOND in predicting the run of circular velocity in galaxies
from the observed distribution of detectable matter is one of the strongest
arguments in its favor.  No simple prescription for reducing the
number of dark halo parameters works as well the MOND formula
(see Sanders and Begeman 1994).  At the very least, the simple MOND formula
provides the most efficient
fitting algorithm for spiral galaxy rotation curves.  This implies that, 
whatever its cause, the discrepancy between the Newtonian dynamical mass
and the visible mass appears below a fixed universal acceleration.

BBS could be, and have been, criticized for being too selective.  One 
might ask, to what extent are criteria applied, at least unconsciously, 
which eliminate those cases which contradict MOND.  Partly to respond
to such criticism and partly because the sample of galaxy rotation 
curves available in the literature is now larger, it was decided to
repeat the analysis of BBS for a less stringently selected sample.
An additional 22 galaxies are considered here, most of which are published
or will soon be published.  Because dark halo models have been presented
elsewhere and because the experience is that multi-parameter dark halo
models can always fit rotation curves, the only comparison is with
the rotation curve predicted by MOND using the observed distribution of
detectable matter in so far as it is traced by the visible light and
the neutral hydrogen.

The essential result of this work is that MOND, with one free parameter
per galaxy-- in some cases two if a bulge is present--, accounts for 
the magnitude of the discrepancy and reproduces
the general shape of the rotation curves of these 22 additional spiral
galaxies.  
This is particularly striking when one keeps in mind the observational
caveats mentioned above and considers that the galaxies
in the present sample have asymptotic rotation velocities ranging
from less than 60 km/s to 300 km/s and cover a range of 1000 in 
luminosity.  In some individual cases
the fits are remarkably good, comparable to the best fits in the 
highly-selected BBS sample.  The values of the fitted parameter,
the luminous mass or implied value of M/L, are reasonable 
in terms of population synthesis models and
have a small scatter
particularly in the near-infrared.   On the whole, the idea is
given further support by comparison with this larger, less stringently
selected sample of galaxy rotation curves.     

\section{The sample}

The 22 galaxies considered here are listed in Table 1.
These combined with the 11 galaxies in BBS comprise the current total 
sample of objects with well-measured 21 cm line rotation curves and with
accurate surface photometry in at least one color band.
There is much overlap with similar lists given by Broeils (1992) and
by Rhee (1996).
Most of the rotation curves here were taken directly from the literature,
but two have not yet been published (M 33, NGC 1003)
and were communicated privately.

This is a motley collection of galaxies observed either at
the VLA or at Westerbork, with varying degrees of precision.
The sample includes
several very large and luminous galaxies such as UGC 2885 and NGC 801,
originally made famous by Rubin et al. (1985) as examples of spiral galaxies
with extended non-declining rotation curves, as well as gas-dominated
dwarfs such as IC 2574 and DDO 168 with gently rising rotation curves.
There are two galaxies with distinctly declining rotation curves,
NGC 2683 and NGC 3521, observed by Casertano and van Gorkom (1991) who
emphasize this general trend in high
surface brightness galaxies. There is one recently-observed  
low surface brightness galaxy, UGC 128 (van der Hulst et al. 1993,
de Blok et al. 1995) and one blue compact galaxy, NGC 2915 (Meurer et al.
1994, 1996).

The columns in the table are generally self-explanatory.  The objects
are listed in order of decreasing asymptotic rotation velocity.
The adopted
distance (column 3) is of critical importance in MOND fits because
the internal accelerations scale inversely as the distance.  Indeed,
as was demonstrated by BBS, the distance can be taken as a free 
parameter of the MOND fit. The Hubble law distance is taken 
for the more distant objects
with ${\rm H_o = 75\,\, km/s/Mpc}$ corrected for local group motion.
Recently determined Cephied distances are taken for M 33 (\cite{mf91})
and NGC 300 (\cite{feta92}).  Distances to other nearby galaxies
(e.g. the Sculptor group galaxies) have been taken from the listed
references.  For several objects a correction for Virgo-centric 
inflow was included following Kran-Korteweg (1986), but in general this
did not differ from the straight Hubble law distance by more than 10\%. 
The corrected blue luminosity (column 4) is taken from the primary 
references and the near-infrared luminosity (column 5) is calculated from 
the H-band apparent magnitudes given by Tormen
and Burstein (1995).  The extent of the observed HI rotation
curve is given in column 6.  The total gas mass,
hydrogen plus helium, assuming that this is 1.3 times the measured
HI mass, is given in column 7.  The value of
the rotation velocity at the outermost radius is given in column 8 
and the corresponding centrepital acceleration, in units of $10^{-8}
{\rm cm/s^2}$, in column 9.  Column 10 gives the numbered
references for the rotation curve and for the photometry.

Nine of the galaxies in this sample show clear evidence in the radial
light distribution for a central bulge component; these are listed in
Table 2.  Because the bulge is
generally assumed to have a more spheroidal shape and because one may wish
to assign a separate mass-to-light ratio to the bulge component, a
decomposition of the profile into bulge and disk components is necessary.
The bulge-disk decomposition given in the indicated references 
is taken in the cases of NGC 5907,
NGC 3521, and NGC 2683.  In the cases of UGC 2885, NGC 801 and NGC 2998
the light distribution for the bulge is taken from the
double exponential decompositions (i.e., exponential bulge and disk) 
by Andredakis and 
Sanders (1994).  Additional double exponential decompositions were 
supplied for NGC 5533, NGC 6674 and NGC 5371 
by Andredakis (private communication).  In most of these cases of an
exponential bulge fit, the radial distribution of disk light was 
determined by subtracting the smooth bulge profile from the total
radial intensity profile (i.e., the disk light distribution is not
assumed to be exponential but is that
actually observed after subtraction of the bulge).
The length scale (the exponential scale length or effective radius) 
and axial ratio of the bulge and 
bulge-to-total luminosity ratios are given in columns 4 and 5 of 
Table 2.

\section{The procedure}

To calculate MOND rotation curves, the same procedure was followed as in
BBS.  The first step is to determine the Newtonian rotation curve of
the detectable matter.  This is done by assuming that the light in
the disk is a precise tracer of the luminous matter (i.e., no radial
variation of M/L in a given galaxy) and that this matter 
has an axially symmetric distribution in an infinitessimally thin disk.
In those
nine cases where there is a bulge-disk decomposition, it is also assumed
that the mean radial distribution of light in the bulge component
traces its luminous mass 
distribution and, for simplicity of calculation, 
that the bulge mass distribution is spherically symmetric.
These calculations were repeated assuming  
that the bulge was highly flattened (a disk), but in general there was
no significant difference in the rotation curve fit (although the fitted
bulge mass is lower).

The gas mass distribution is assumed to be traced by the mean radial
distribution of neutral hydrogen.  The HI surface density is everywhere
increased by a factor of 1.3 to account for the contribution of helium.
This, of course, neglects any contribution of molecular gas, which, in
the absence of more detailed information, is assumed to be distributed as
the luminous component (Young 1987).  The gas distribution is also taken to
be axisymmetric in an infinitessimally thin disk.

Given the Newtonian acceleration, ${\bf g_n}$, the true gravitational
acceleration ${\bf g}$ is determined from the MOND formula
$$ \mu (g/a_o) {\bf g} = {\bf g_n} \eqno(1) $$
where ${\rm a_o}$ is the MOND acceleration parameter and 
$$\mu(x) = x(1+x^2)^{-1/2}. \eqno(2) $$
This commonly assumed form has the appropriate asymptotic behavior
yielding Newtonian dynamics in the high acceleration limit and MOND dynamics
in the low acceleration limit (Milgrom 1983).  For several galaxies in
the sample (e.g. IC 2574), $g<<a_o$ everywhere, so the exact form of
$\mu$ is unimportant.  The rotation law is given, as usual, by
$$ {{v^2}\over r}   =  g \eqno(3) $$ 
which means that, with eqs.\ 1 and 2, as r becomes large
$$v^4 = GM_ta_o \eqno(4)$$
where ${\rm M_t}$ is the total finite mass of the galaxy in
stars and gas ($M_t= M^* + M_g$) .
In the determination of ${g}$ it would be desirable to apply the
physically consistent field equation of Bekenstein and Milgrom (1984), 
but this is computationally difficult.  Moreover, it has been demonstrated
that the simple MOND formula gives, in most cases, results which agree quite
closely with the integration of the full field equation (Milgrom 1986,
Brada and Milgrom 1995).  In the context of inertia-modified
theories of MOND (\cite{ml94}), eq.\ 1 would be exact. 

The observed rotation curve is fit in a least-square program applying
eqs. 1, 2, and 3.  The free parameter of the fit is always ${\rm M_d}$, 
the total mass
of the luminous disk, and for those cases in Table 2, ${\rm M_b}$, 
the mass of the bulge.  
In combination with the observed luminosities this yields
the mass-to-light ratio of the luminous components.    
The total stellar mass of a galaxy is $M^* = M_d + M_b$.

Here, the acceleration parameter ${\rm a_o}$
is not allowed to be free but is taken to be the mean value determined
by the fits to the higher-quality rotation curves of BBS; i.e., 
$1.2\times 10^{-8}\,\, ({\rm H_o/75\,kms^{-1}Mpc^{-1}})\,\, {\rm cm/s^2}$.  
It is a questionable procedure,
in principle, to take a quantity supposed to be a fundamental constant as
a fitting parameter (\cite{ml88}).
Moreover, unlike BBS who fit the
rotation curves allowing the distance to a galaxy to be both fixed and
free, here we fix the distance at the adopted value given in
Table 1.  This is done because it is desirable to reduce the dimensionality
of the parameter space when the data are less precise.
Random or systematic errors in the estimated circular
velocity at a few points can yield solutions in an extreme region of 
a multi-dimensional parameter space; i.e., minima in the $\chi^2$ 
surface which appear to be sharp are, in fact, very broad due
to under-estimated errors in observed rotation velocity.
In principle, this problem could be 
eliminated by a realistic estimate of the errors, but in practice this is
very difficult when there are unknown systematic effects (i.e., warps, bars,
pressure support, beam smearing).  Therefore, with ${\rm a_o}$ and distance
fixed, the quality of the fit can be judged by visual inspection
and the plausibility of the implied M/L ratios.

\section{Results}

The results are given in Fig.\ 1 and in Table 3.  In the figure we see
the observed rotation curve (points with error bars) compared with the
fitted MOND rotation curve (solid line).  The Newtonian rotation curves
of the various individual components are also shown as explained in
the caption.  

A word is necessary about the indicated error bars.  These are
taken, in general, directly from the reference for the rotation curve
(Table 1) and cannot be interpreted in a uniform way.  Often the indicated
errors are formal one-sigma errors returned by the program which fits
tilted rings to the two-dimensional HI velocity field.  These are 
unrealistically low because this technique does not include an 
assessment of the possible systematic effects.  In other cases, error bars
are estimated by performing the tilted ring analysis separately for
two different sides of the galaxy (approaching and receding) and taking
the difference between the resulting rotation curves.  This gives a
fairer assessment of those systematic errors resulting from asymmetries in the
velocity field.  In any case, the commonly used tilted-ring algorithm
is, at best, a first order correction to the effects of warping in estimating
the circular velocity.

The rotation curves are given as listed in
Table 1, ordered in decreasing asymptotic rotation velocity.
The first rotation curves are those of large luminous systems with 
bulges or at least central light concentrations.  These are distant
galaxies so the spatial resolution of the 21 cm line observations is
generally several kiloparsecs.  The MOND fits indicate
that the mass distribution in several of these galaxies 
is more centrally concentrated than the
light distribution implying that the bulges have a significantly higher
M/L than the disk.
The rotation curves in the last two panels are those
of relatively nearby dwarfs.  These are systems without bulges where
the gas makes a significant contribution to the total Newtonian force
in the outer regions.  Several of these systems are irregular with
asymmetric velocity fields.  

It is necessary to discuss several of the individual galaxies in greater 
detail:

{\it NGC 5533}:  This large Sab galaxy is the earliest type in the sample.  
The HI surface densities are rather low and the distribution in the
outer parts is quite patchy.  There are also significant side-to-side
asymmetries in the outer velocity field as well as kinematic evidence
for a warp (Broeils 1992).  Because of this and the low spatial resolution
the galaxy would not meet the BBS criteria.
The double exponential bulge-disk decomposition is provided by 
Andredakis (private communication) and implies that a large fraction of
the total light is in the bulge.  However, the MOND fit to the rotation
curve requires that an even larger fraction of the mass is in the bulge.
This leads to a bulge M/L of seven and a disk M/L of about one (both in the
blue band).  While the overall mass-to-light of three in the
blue implies that MOND successfully accounts for the magnitude of the 
discrepancy in this galaxy, near-infrared photometry would be considerable
interest in this case in estimating the distribution of the stellar
mass in the central regions.
 
{\it NGC 6674}:  The global blue M/L of 2.6 implies that MOND successfully
accounts for the magnitude of the discrepancy;  however, the detailed fit
is the worst of the sample. The large central rotation velocities and 
mildly declining rotation curve require a strong central mass concentration.   
The double exponential
decomposition of the radial light profile by Anderedakis (private 
communication) does imply that a large fraction of the total 
luminosity is in the bulge, and this yields reasonble mass-to-light ratios 
for the bulge and disk.  But the real difficulty with the use of this 
rotation curve is that the galaxy is conspicuously barred (Broeils and
Knapen 1991).  Moreover the bar is oriented along the apparent minor
axis of the galaxy as projected onto the sky which, because of elliptical
streaming, would have the effect of increasing the apparent rotational 
velocity in the inner regions.  Because of the large-scale non-axisymmetric
structure this galaxy is clearly not very suitable for detailed rotation
curve modelling.

{\it NGC 5907}:  This large, relatively nearby edge-on galaxy has a 
well-determined and very extended HI rotation curve (Sancisi and van Albada
1986).  However, it had not been possible to estimate the Newtonian
rotation curve due to the luminous matter because the high dust obscuration
in the plane of the galaxy masks the true radial light distribution.
This has changed with the near-infrared photometry of Barnaby and
Thronson (1992, 1994) which indicates a more centrally concentrated
distribution of lumionous material.  This highlights the value of 
near-infrared photometry as
the most accurate, absorbtion-free tracer of the dominant stellar component.
Here, the decomposition by Barnaby and Thronson 
into an exponential disk and a bulge represented
by a modified Hubble profile is used directly to calculate the Newtonian
rotation curve; i.e., because the galaxy is edge-on the exponential model
for the disk is used rather than the detailed photometry.

{\it NGC 3521 \& NGC 2683}:  These are given by Casertano and
van Gorkom (1991) as examples of galaxies with declining rotation curves.
The contribution of the gas to the Newtonian rotation curve is not
shown here because the mean radial distribution of the gas is not
given in this reference; in any case, the total gas mass, estimated from
global 21 cm line profiles (Table 1), is less than  
10\% of the fitted disk mass (Table 3) in both cases.
The photometry by Kent (1985) includes decompositions into exponential
disks and ${\rm r^{1/4}}$ law bulges.  Here the models rather than the 
detailed light distributions are used in determining the Newtonian 
rotation curves.  The observed rotation curves are not ideal for
estimating the true run of circular velocity. In neither case is the
rotation curve determined from a full two-dimensional 
radial velocity field.  In NGC 3521,
the distribution of HI is asymmetric, extending 20\% further on one
side than on the other, and the velocity structure is asymmetric;
NGC 2683 is near edge-on.  Thus these galaxies fail in several respects
to satisfy the selection criteria of BBS.  Nonetheless, 
the rotation curve fits demonstrate that MOND is quite capable
of reproducing declining rotation curves if the mass distribution is
sufficiently centrally concentrated, a point made by Milgrom in his 
original papers (1983).  In NGC 3521, the abrupt decline in rotation velocity
between of 20 kpc and 28 kpc could be, if confirmed, problematic for MOND,
although it would also be problematic for Newtonian dynamics 
since the decline is steeper than Keplerian.

{\it UGC 128}:  This is a low-surface-brightness (LSB) spiral with
an extrapolated B-band central surface brightness fainter than
23 ${\rm mag/arcsec^2}$ (de Blok et al. 1995).  Although it is faint,
the linear size is large with the HI rotation curve extending to 40 kpc.
Because the implied surface density is below the MOND critical 
surface density of $a_o/G$, the MOND prediction (Milgrom 1983) is that
the discrepancy should be large within the optical disk and that
the rotation curve should be slowly rising to its asymptotic limit.  This
is seen to be the case and the MOND rotation curve agrees with the
observed curve in detail.  It should be emphasized that the general
MOND prediction of a large discrepancy in low surface density
systems (Milgrom 1983) 
was made long before observations of systems such as this one confirmed it.

{\it M33}:  The observed rotation curve of this classic nearby Sc spiral
is from an unpublished analysis by Kolkman (1995) based upon observations of
Deul and van der Hulst (1987, see also \cite{mhr96}).  
The large number of independent observed points on the rotation
curve, due to
the large angular size this object, and the well-established
Cepheid distance (Madore and Freedman 1991), make this a good case
for detailed rotation curve fitting; although, there is a 
significant warp in the outer regions.  

{\it NGC 2915}:  This is a blue compact galaxy (BCG) recently analyzed
by Meurer et al. (1996).  The neutral hydrogen extends well beyond the bright
optical image (to 22 times the exponential scale length), and the
observed rotation curve remains constant with a suggestion of a rise
at the outermost measured points.  This implies a very large discrepancy
between the visible and Newtonian dynamical mass 
(hence, the authors refer to this object as the darkest disk galaxy).
The MOND rotation curve is higher than the observed curve in the inner
regions (where the errors are large) but agrees very well with the 
observed rotation curve in the outer regions.  Here even the apparant
rise of rotation velocity in the last few points is reproduced due to the
contribution of the gas to the dynamical mass.  However, in spite of this
agreement, the implied mass-to-blue light ratio for the disk is 6.9 which is
an uncomfortably large value for a BCG.  
The distance to the galaxy is quite uncertain;
Meurer et al. give $5.3 \pm 1.3$ Mpc based on the method of brightest 
stars.  At a distance of 6.6 Mpc the M/L is reduced to 3.3.
It might also be that the luminosity has been underestimated if a 
faint luminous halo surrounds the bright blue compact object.

{\it DDO 168}:  This is a dwarf with a small bar in the central regions.
The MOND rotation curve lies noticably above the observed rotation
curve in the inner regions.  This should not be given too much significance
because of the possible effects of the bar or of beam smearing.

Table 3 lists the fitted disk and bulge masses for all galaxies in the
present sample as well as in the sample of BBS-- a total of
33 galaxies for which MOND rotation curves have been calculated.  
Also shown are the
implied mass-to-blue light ratios, for the disk and bulge separately where
applicable.  In column 6 the the global (bulge plus disk) mass-to-light 
ratio in the blue is given for the luminous (stellar) component, ${\rm 
M^*/L_B}$;  in column 7 the ratio of the total
mass (stars plus gas) to luminosity in the blue band is given
(${\rm M_t/L_B}$);  in column
8 the ratio of the total mass to luminosity in the H-band (${\rm M_t/L_H}$) 
is given for 
those galaxies for which an H-band magnitude has been measured.
It should be noted that the fitted mass, $\rm M^*$, includes not only
the mass in luminous stars but also any other component which is 
distributed like the stars, such as, possibly, the molecular gas.

We see that in most cases the global mass-to-light ratios are 
reasonable and consistent
with population synthesis models;  i.e., the models imply
blue-band M/L values in the range from a few tenths to 10 depending upon
the star-formation history and metallicities ({\cite{bc93}, \cite{wo94}).  
More importantly, there are no very high
values (the highest being for NGC 2915 discussed above) which means that
MOND can certainly account for the magnitude of the global mass discrepancy;
i.e., there is no suggestion that additional unseen matter is needed.
Although there are several very low values of ${\rm M^*/L_B}$ (see below), 
none of the fitted masses is negative (this is possible considering that
the gas mass is measured directly); 
i.e., in no case does MOND seriously overcorrect for the discrepancy. 

\section{Global mass-to-light ratios and the Tully-Fisher relation}

Fig.\ 2 shows the global blue stellar mass-to-light ratios (${\rm M^*/L_B}$)
of the sample 
galaxies determined from the MOND fits (column 6 of Table 3) plotted as 
a function of asymptotic rotation velocity.  There is a range of a factor of 
almost 100, but there is also an apparent trend
of increasing ${\rm M^*/L_B}$ with increasing rotation velocity.  Due to
the sensitivity of the blue luminosity to star-formation activity, this
would be consistent with more active star formation in the lower luminosity
gas-rich galaxies in this sample.  But the overall consistency of the 
implied ${\rm M^*/L_B}$ values with population synthesis models is 
demonstrated by Fig.\ 3.  This is a plot of the MOND ${\rm M^*/L_B}$ vs.
B-V color for those galaxies in the sample with a reddening-corrected
B-V listed in the Third Reference Catalogue (de Vaucoleurs et al. 1991).  
Also shown is the ${\rm M/L_B}$ 
as a function of B-V predicted from the population synthesis
models of Larsen and Tinsley (1978);  here, the properties are those
of a population of stars evolved for $10^{10}$ years with various 
prescriptions for a monotonically decreasing star formation rate.  It is
seen that the general trend of decreasing M/L with increasing blueness
is present in the MOND values of ${\rm M^*/L_B}$.

There are, however, five galaxies in Fig.\ 3 with implied values of 
${\rm M^*/L_B}$ less than 0.25:  NGC 55, DDO 168, NGC 1003, IC 2574, 
and DDO 154.  While such low mass-to-light ratios are possible in extreme 
starburst galaxies, it should be noted that all of these objects are
nearby ($<$ 4 Mpc) gas-rich dwarfs.  For such objects 
the implied ${\rm M^*/L}$ values are
extremely sensitive to the adopted distance.  For example, for NGC 3109
(not plotted here), the distance used by BBS is a Cephied-based
estimate of 1.7 Mpc (Sandage and Carlson 1988).  At this distance, 
the measured 
mass of gas is almost 100\% of the MOND mass (eq.\ 4) which means that
that ${\rm M^*/L_B}\approx 0$.  A more recent Cephied distance estamate
is 1.3 Mpc (Cappacioli et al. 1992).  At this distance the gas mass is reduced
to about 60\% of the MOND mass which means that ${\rm M^*/L_B}$ is increased
to 0.6.  So given the distance uncertainties in these gas-rich dwarfs,
it is not surprising that some of the fitted values of the stellar M/L
would be unrealistic;  the point is that ${\rm M^*/L_B}$ is the single
fitted parameter and must reflect all uncertainties involved in this
procedure. 

Figs.\ 4 and 5 show the observed B- and H-band luminosity-rotation
velocity relationships (Tully-Fisher) for the galaxies in the combined 
total sample (Table 3).  Here the rotation velocity is that measured at the
most distant points for which the determination is reliable;  this would
correspond most closely to the asymptotic circular velocity in the
context of MOND (eq.\ 4).  The H-band relation is plotted for those
15 galaxies in the combined sample with measured H-band magnitudes (Tormen
\& Burstein 1995).  In both cases, the relation appears quite linear on the 
log-log plot with a slope near the canonical value of four.  The slopes
are somewhat larger (4.0 $\pm 0.25$ in the blue, 4.4 $\pm 0.19$ 
in the near-infrared) 
than usually encountered primarily because the use of the actual 
rotation curve gives a larger value of the rotation velocity
for the low luminosity galaxies than does the global 21 cm line profile.
The scatter 
in log luminosity is 0.30 in the blue corresponding to 0.75 magnitudes.
In the near-infrared the correlation is much tighter (as is well-known)
with a scatter of 0.12 (0.3 magnitudes).  The tightness of the relation
implies that the errors in distance (at least for this subsample of
15) are not large.

In the context of MOND, there is a total mass-asymptotic rotation velocity
relation which is {\it exact} (eq.\ 4).  There is a similar
luminosity-velocity  relation only for a luminosity indicator which
is proportional to the mass, i.e.,
$$ v^4 = Ga_o (M_t/L) L \eqno(5)$$
  Therefore, the scatter in the observed
relation, apart from observational errors (e.g., inaccuracies in the
rotation velocity or the global magnitude, errors in the distance),
would only be due to intrinsic scatter in the mass-to-light ratio.
The tightness of the observed relation suggests that this might,
in fact, be small.

However, as we see in eq.\ 5,
it is not the mass-to-light ratio of the luminous matter ${\rm M^*/L}$
which is relevant to the slope and scatter in the observed
TF relation but rather the {\it total mass}-to-light ratio (${\rm M_t/L}$).
The gas mass can make
a very significant contribution to this total 
in the low-mass, low luminosity galaxies. 
In Fig.\ 6 we see the total mass-to-blue light ratio (column 7 of
Table 3) plotted against the asymptotic rotation velocity, where now the
total mass includes the observed neutral hydrogen plus implied helium mass.  
The range in this quantity is about a factor of 15;  the mean
value is 1.9 with a dispersion of 1.7;  
that is, the scatter is now about 90\% which is
quite consistent with the 97\% scatter observed in the blue TF relation
(Fig.\ 4). 
It is also evident that for a number of the low-luminosity dwarf galaxies, 
the total mass-to-light ratio is quite large, opposite to the trend noted
in Fig.\ 2.  This is entirely due to the large contribution of
the (non-luminous) gas to the total mass in these systems.  This increase
in actual mass-to-light ratio can result in a steepening of the observed
TF law at the low-luminosities-- a steepening which has been previously
noted in much larger samples (\cite{aaeta82}).

In Fig.\ 7 we see the total mass-to-light ratio in the near-infrared 
(column 8 of Table 3) plotted 
against rotation velocity.  Here the range in
M/L is reduced to about a factor of two;  the mean value is
2.3 with a dispersion of 0.73.  
This scatter of 31\% is again entirely consistent with
the scatter in the observed infrared TF law of 33\% (Fig.\ 5).  
There is also a
slight trend of increasing ${\rm M_t/L}$ with decreasing rotation velocity.  
This 
again is due to the increasing contribution of the gas mass in the 
lower luminosity systems 
and would be consistent with a slope somewhat larger than four in the 
observed TF relation. 

\section{Is a definitive falsification possible?}

The one-parameter MOND rotation curve fits to the galaxies in the 
highly selected sample of BBS are, with one exception, in very precise
agreement with the observed rotation curves.  In the larger sample 
considered here a number of the rotation curves are not perfectly fit;
one can argue that this is because of the
various uncertainties mentioned (distance errors, beam smearing,
warps, non-circular motions, contribution
of molecular gas, inprecise bulge-disk decomposition, the use of 
visual rather than infrared magnitudes to trace the stellar density
distribution, true radial variations
in M/L of the stellar population).  None-the-less, the overall
form of the rotation curves and general trend with luminosity 
is generally quite well reproduced, with the
MOND rotation curves of luminous high surface brightness galaxies
exhibiting the rapid rise and then decline to an asymptotic value and
those of the low luminosity, low surface brightness galaxies rising
slowly to the asymptotic value as is observed.  But  because of the
unknown systematic effects which may give rise to a difference between
the measured rotation curve and the true run of circular velocity, it
is not useful to apply a statistical goodness-of-fit 
criterion to assess objectively the success of MOND in reproducing
observed rotation curves from the distribution of detectable matter.

Then the question naturally arises
of what would constitute a bad MOND rotation curve fit.  Is it in fact
possible to definitively falsfy MOND by this technique?  Can an example
be given where MOND fails fundamentally to predict the observed rotation
curve of a spiral galaxy?  There is such a case, and that is the
exceptionally bad fit to NGC 2841 in the sample of BBS if this galaxy is
at its Hubble law distance of 9.5 Mpc (h=0.75).  
This MOND rotation curve fit to NGC 2841 is
reproduced in Fig. 8a.   Not only is the rotation curve badly fit but the
implied bulge and disk mass-to-light ratios are outrageous: $(M/L)_b
= 0.6$, $(M/L)_d = 13$.  The basic problem is that the form
of the observed rotation curve compared to the Newtonian rotation curve
of the detectable matter, suggests that a large discrepancy is present 
at accelerations larger than $a_o$ which is not possible in the context of
MOND.  The data on which the measured rotation curve is based are of
the highest quality in the existing literature (Begeman 1987):  there 
is sufficient spatial resolution; the HI distribution is reasonably 
smooth and symmetric; the outer warp evident in the gas kinematics  
is symmetric and well-modelled by the tilted ring algorithm.
Only a large and improbable positive radial gradient in M/L in the 
disk itself could allow this rotation curve to be explained by MOND 
using the standard value of $a_o$.  

BBS noted that if the distance is allowed to be a parameter in the 
least-square fit, MOND fits for most of their sample improve
slightly;  the fitted distance agrees well with the Tully-Fisher distance
and is generally within 15\% of the
Hubble law distance.  The one exception is the case of NGC 2841 
which requires a 
distance twice as large as the Hubble law distance to achieve a reasonable
MOND fit; i.e. 19.3 Mpc rather than 9.5 Mpc (the luminosities and M/L
values for this object
given in Table 3 are based upon this larger distance).  The MOND rotation
curve fit to this galaxy at the larger distance is shown in Fig.\ 8b.  Here
the fitted curve agrees well with the observed curve and the implied
M/L values are much more reasonable:  3.5 for the disk and 4.3 for the bulge
(based upon a new double-exponential decomposition by Andredakis).
NGC 2841 is the only galaxy out of the total sample of 33 which
requires a distance substantially different than the Hubble law distance
in order to achieve a reasonable MOND fit to the rotation curve.
This larger distance is consistent with all Tully-Fisher 
determinations; for example, Aaronson and Mould (1983) give a 
distance of 15.6 Mpc to the NGC 2841 group based upon the H-band 
Tully-Fisher relation.  But in fact, Tully-Fisher distances are not
independent of the fitted MOND distance because MOND subsumes the Tully-Fisher
relation (eq.\ 5).  Thus a truly independent and reliable
distance estimate to this
galaxy (e.g., Cephieds) offers the possibility of a definitive 
falsification of MOND:
if the galaxy is close to its Hubble law distance, the viability
of MOND is seriously threatened (one well-established
counter-example is sufficient);  if the galaxy is twice as far away
as the Hubble law distance, MOND remains viable.

Given the existence of large scale flows, it is not surprising that one
galaxy out of 33 might have a distance significantly different from that
implied by uniform Hubble flow; however, there is a Hubble law and
it would be quite negative for MOND if distances to
several galaxies in the sample had to be adjusted by such a large factor
in order to achieve reasonable fits.

\section{Conclusions}

The 22 galaxies considered here along with the 11 galaxies previously
considered by BBS comprise the current total published 
sample of galaxies (plus or minus two or three) 
with optical or infrared surface photometry and with observed HI rotation 
curves extending well beyond the optical image of the galaxy.  Although
this number will rapidly grow due to several large surveys now underway,
this sample of 33 galaxies constitutes, at present, 
the entire body of data relevant to the nature 
of the discrepancy between the classical dynamical and visible mass in 
galaxies (the several hundred optical rotation curves in the literature
do not, in general, extend far enough to probe the systematics of
the discrepancy).  BBS considered a highly selected sub-sample of this
collection of rotation curves-- galaxies for which one can be reasonably 
sure that the measured 21 cm line rotation curve gives a fairly good
estimate of the run of circular velocity and thus the radial force
beyond the optical image.  They demonstrated that the observed
distribution of detectable matter in the galaxies, in the context of MOND, 
reproduced the observed rotation curves quite accurately, often down
to rather small details, without invoking unseen matter.

Here the remainder of the current total sample of these galaxies, 
i.e., objects which either do not meet the selection criteria of BBS
or those added since 1991, has been considered in the context
of MOND.  Compared to the BBS sample, some deterioration 
in the quality of the fits is expected and seen; but
in general, the form and amplitude of the observed rotation curves is also
reproduced for these addtitional galaxies.  The reader, when assessing
the quality of the fits in Fig.\ 1, should keep in mind that, 
unlike the usual
dark halo models, there is only one, or in some cases two, 
adjustable parameters per galaxy and
that is the mass of the luminous components.  The MOND acceleration parameter
has been fixed at the BBS value of $1.2\times 10^{-8}$ ${\rm cm/s^2}$ 
(normalized to the distance scale of ${\rm H_o} = $ 75 km/s Mpc), and
the distance to any galaxy has been fixed at its value determined via
the Hubble law or by more direct methods (in particular Tully-Fisher 
distances have not been used since these are, in effect, the MOND distances).

For the large luminous galaxies with central bulges or light concentrations,
such as NGC 5533 and NGC 801, the MOND fits require a mass 
distribution which is even more centrally concentrated than the light
distribution.  This implies (see Table 3) a
bulge M/L which is significantly larger than the disk
M/L, but the exact values depend quite critically upon
how the light distribution is decomposed into bulge and disk contributions.  
Because of this complication of bulge-disk decomposition and the
possibility of an extra free parameter, these systems, with respect
to rotation curve analysis, are not as clean as pure disk galaxies.
But in general a higher bulge M/L would be more consistent with
expectations for possibly older, or at least less actively 
star-forming, spheroidal sub-system.  It is significant that for those
galaxies without a conspicuous bulge MOND does not require, 
in any single case, a larger central M/L in order to achieve 
a reasonable fit to the rotation curve; that is to say, the necessity
of a larger central M/L occurs only in those cases where the obvious
presence of a separate bulge component justifies it.

MOND does quite well in reproducing the rotation curves of the 
pure disk systems (e.g. M 33); in particular, the scheme works well for the
low surface-brightness galaxy, UGC 128, and for the gas rich
systems such as NGC 2915, NGC 55 and IC 2574.  This not only lends support
to MOND but also to the assumptions which underly the whole procedure,
such as constancy of the mass-to-light ratio within the disk component of
any given galaxy and
the absence of a significant contribution to the detectable mass
by molecular gas with a radial density distribution differing from
that of the luminous disk.

When assessing the quality of MOND or dark halo fits to galaxy rotation
curves, one should also consider the physical plausiblity of the fitted 
parameter(s).  In this work, if we neglect the ambiguous procedure
of bulge-disk decomposition, the only parameter to consider is
the global stellar mass-to-light ratio.  We have noted above that the 
implied global ${\rm M^*/L_B}$ values, although 
spanning range from 0.1 to 7, are not so high as to require 
substantial additional dark matter, which would
be contrary to the spirit of MOND, nor are there negative values  
which would suggest that MOND substantially overcorrects
(a number of negative stellar mass-to-light ratios would constitute a 
falsification
of the theory).  Moreover, the range in this one fitted parameter must 
reflect all of the uncertainties of the procedure:  distance errors,
non-circular motion, the use of visual photometry, etc.  Even so,
the fitted values of ${\rm M^*/L}$ in the blue and H-band
are generally consistent with 
those implied by stellar population models and exhibit the trend of
increasing ${\rm M^*/L_B}$ with redder color. 

When the total mass (the fitted luminous mass plus measured gas
mass)-to-light ratio is considered, the scatter is reduced considerably and
becomes quite consitent with the observed scatter in the blue Tully-Fisher
relation. This is even more 
striking in the near-infrared, where the scatter in the total 
mass-to-light ratio is reduced to the order of 30\% which again is comparable
to the scatter in the near-infrared Tully-Fisher relation.
Not all of this scatter is intrinsic; certainly some  
of the apparent scatter in ${\rm M_t/L}$ results from errors in the estimated 
distances (in MOND, the total mass estimate is fairly independent of distance
implying that ${\rm M_t/L}$ scales as the inverse square of the distance).  
With this in mind, the overall
small scatter in the implied ${\rm M_t/L}$ values, 
where ${\rm M_t}$ includes the fitted MOND
mass for the luminous component, certainly argues forcefully for the 
plausibility of the implied MOND masses.

For the combined sample of 33 galaxies, MOND fails only in one case:
the MOND fit to the well-determined rotation curve of NGC 2841 is not 
acceptable.  The failure is serious; not only is the form of the observed
rotation curve not reproduced but implied mass-to-light ratios of the
bulge and disk are implausible (the implied disk M/L of 13 means that
MOND fails to account for the magnitude of the discrepancy).  If, however,
NGC 2841 is twice the distance implied by the Hubble law, the the predicted
MOND curve agrees with the observed curve in detail and the implied
M/L values are reasonable.  An independent distance determination to
this galaxy (i.e., independent of the Tully-Fisher relation) is therefore
crucial for MOND;  a distance significantly less than 19 Mpc would falsify
the idea. 
 
In general, the analysis of this 
larger sample reinforces the conclusions of BBS:  in terms of the number
of parameters
MOND provides the most efficient description of the systematics of galaxy
rotation curves.  Given the observed distribution of light and gas in
a galaxy, one may predict with considerable precision the extended rotation
curve which is actually observed
by adjusting only the M/L of the luminous component, and this required
M/L (in the near-infrared at least) lies within rather small range 
around a mean value of approximately two in solar units. 
But it is not just that MOND requires a smaller number of parameters 
to describe rotation curves than do dark halo models.  The philosophy
of rotation curve fitting
is really quite different for the two hypotheses: dark matter rotation
curves are fits which define the properties of the dark halo; 
MOND rotation curves are predictions which test the validity of the theory.
MOND is a viable
alternative to the dark matter hypothesis precisely because of its
predictive successes.

Because of surveys presently underway, there will soon
be a rapid increase in the number of high-quality rotation curves in
the literature.  In the context of the ``missing mass problem'', it
will be of great interest to assess the continued performance of MOND
in predicting the rotation curves for a larger number of spiral galaxies
with a wider range of properties.

\acknowledgements

I am very grateful to a number of people for either sharing their data
with me in advance of publication or for providing existing
data in convenient form.  This includes H. Hoekstra, M.-H. Rhee,
E. de Blok, R. Sancisi and A. Broeils.  In particular, Adrick Broeils 
has done this analysis for a number of these galaxies observed as part
of his Ph.D. dissertation, and he has recently 
confirmed several of the cases shown here.  I thank G.R. Meurer
and C. Carignan for data in advance of publication and for very useful 
comments on their observations of the blue compact galaxy NGC 2915.  
I thank Y. Andredakis for his careful bulge-disk decompositions.
As always, when it comes to 
analysis of rotation curves, the advice and help of K.G. Begeman is
invaluable.  I thank him especially for initiating me into the wonders
of GIPSY.  And finally, I am most grateful to M. Milgrom. 
His comments and insight  
have been, as always, an enormous help and encouragement in this work.

\clearpage
\begin{table}[p]
\tighten
\begin{flushleft}
\caption{The sample galaxies listed in order of decreasing rotation 
velocity \label{t1}}
\begin{tabular}{|c|c|c|c|c|c|c|c|c|c|}
\tableline
${\rm Galaxy }$ &${\rm Type}$ & ${\rm D}$ & ${\rm L_B}$ & ${\rm L_H}$
   & ${\rm R_{HI}}$ & ${\rm M_{gas}}$ & ${\rm V_{rot}}$ & ${\rm a}$ & 
   $ {\rm Ref.}$ \\
$  $ & $ $ & $ {\rm Mpc} $ & ${\rm 10^{10}\,L_\odot}$ & 
   ${\rm 10^{10}\,L_\odot}$ & ${\rm kpc}$ & ${\rm 10^{10}\, M_\odot}$
   & ${\rm km/s}$ & ${\rm 10^{-8}\, cm/s^2}$ & $ $ \\
 (1) & (2) & (3) & (4) & (5) & (6) & (7) & (8) & (9) & (10) \\ 
\tableline
  UGC 2885 & Sbc & 79 & 21. &   & 73 & 5.0 & 300 & 0.40 & 1,2 \\  
  NGC 5533 & Sab & 54 & 5.6 &  & 74 & 3.0 & 250 & 0.27 & 3,4,5 \\
  NGC 6674 & SBb & 49 & 6.8 & & 69 & 3.9 & 242 & 0.28 & 3,4 \\
  NGC 5907 & Sc & 11 & 2.4 & 4.9 & 32 & 1.1 & 214 & 0.46 & 6,7 \\
  NGC 2998 & SBc & 67 & 9.0 & & 47 & 3.0 & 213 & 0.31 & 3,4 \\ 
  NGC 801 & Sc & 80 & 7.4 &  & 59 & 2.9 & 208 & 0.24 & 1,3 \\
  NGC 5371 & S(B)b & 34 & 7.4 & & 40 & 1.0 & 208 & 0.35 & 8,9 \\
  NGC 5033 & Sc & 11.9 & 1.9 & 3.9 & 35 & 0.93 & 195 & 0.35 & 8,10 \\
  NGC 3521 & Sbc & 8.9 & 2.4 &  & 28 & 0.63  & 175 & 0.35 & 11,12 \\
  NGC 2683 & Sb & 5.1 & 0.6 &  & 18 & 0.05  & 155 & 0.43 & 11,12 \\
  NGC 6946 & SABcd & 10.1 & 5.3 & & 30 & 2.7 & 160 & 0.28 & 12 \\
  UGC 128 & LSB & 56.4 & .52 &  & 40 & 0.91 & 130 & 0.14 & 13,14 \\
  NGC 1003 & Scd & 11.8 & 1.5 & 0.45 & 33 & 0.82 & 110 & 0.12 & 3,4 \\
  NGC 247 & SBc & 2.8 & .35 & 0.22 & 11 & 0.13 & 107 & 0.34 & 15,16 \\
   M 33 & Sc & 0.84 & 0.74 & 0.43 & 8.3 & 0.13 & 107 & 0.45 & 17,18,19 \\
  NGC 7793 & Scd & 3.1 & .34 & 0.17 & 6.7 & 0.096 & 100 & 0.48 & 16,20 \\
  NGC 300 & Sc & 2.15 & 0.3 & & 12.7 & 0.13 & 90 & 0.21 & 15,16 \\
  NGC 5585 & SBcd & 7.6 & 0.24 & 0.14 & 12 & 0.25 & 90 & 0.22 & 21 \\
  NGC 2915 & BCD & 5.6 & 0.036 &  & 15 & 0.1 & 90 & 0.17 & 22,23 \\
  NGC 55 & SBm & 1.6 & 0.43 &  & 9 & 0.13 & 86 & 0.27 & 24 \\
  IC 2574 & SBm & 3.0 & 0.08 & 0.022 & 8 & 0.067 & 66 & 0.18 & 25 \\
  DDO 168 & Irr & 3.8 & 0.022 & & 3.7 & .032 & 54 & 0.26 & 3 \\
\tableline
\end{tabular}
\tablerefs{1, Kent 1986; 2, Roelfsema \& Allen 1985; 3, Broeils 1992b;
4, Broeils \& Knappen 1991; 5, Kent 1984;  
6, Barnaby \& Thronson 1992, 1994; 7, Sancisi \& van Albada 1987; 
8, Begeman 1987;  
9, Wevers 1984; 10, Kent 1985; 11, Casertano \& van Gorkom 1991;
12, Carignan et al. 1990; 13, van der Hulst et al. 1993; 14, de Blok
et al. 1995; 15, Carignan \& Puche 1990b; 16, Carignan 1985;  
17, Deul \& van der Hulst 1987; 18, Rhee 1996; 19, Kent 1987;
20, Carignan \& Puche 1990a; 21, Cote et al. 1991; 22, Meurer et al. 1994;
23, Meurer et al. 1996; 24, Puche et al. 1991; 25, Martimbeau \& Carignan 
1994}
\end{flushleft}
\end{table}
\clearpage
\begin{table}[p]
\begin{flushleft}
\caption{Bulge-disk decompositions \label{t2}}  
\begin{tabular}{|c|c|c|c|c|c|} 
\hline
${\rm Galaxy }$ & ${\rm Bulge}$ & ${\rm r_B}$ & ${\rm b/a}$ & ${\rm B/T}$ &
       ${\rm Ref.}$ \\
$  $ & ${\rm model}$ & $ {\rm kpc} $ & $  $ &  $  $ & $  $ \\
(1) & (2) & (3) & (4) & (5) & (6) \\ 
\hline
  UGC 2885 & exp & 0.6 & 0.86 & 0.07 & 1,2  \\  
  NGC 5533 & exp & 1.7 & 1.00 & 0.42  & 3,4 \\
  NGC 6674 & exp & 1.1 & 0.90 & 0.79 & 3,4 \\
  NGC 5907 & Hub & 0.2 & 0.45  & 0.17 & 5  \\
  NGC 2998 & exp & 1.0 & 0.49 & 0.10 & 1,2  \\ 
  NGC 801 & exp  & 1.1  & 0.75  & 0.35 & 1,2  \\
  NGC 5371 & exp & 0.9  & 1.00  & 0.36  & 3,6  \\
  NGC 3521 & $r^{1\over 4}$ & 0.5  & 0.52  & 0.17 & 7  \\
  NGC 2683 & $r^{1\over 4}$ & 1.7  & 0.21 & 0.30  & 7  \\
\hline
\end{tabular}
\end{flushleft}
\tablerefs{1, Kent 1985; 2, Andredakis \& Sanders 1994; 3, Andredakis,
private communication 1996; 4, Broeils and Knapen 1991; 
5, Barnaby \& Thronson 1992, 1994; 6, Begeman 1987; 7, Kent 1985} 
\end{table}
\clearpage
\begin{table}[p]
\begin{flushleft}
\caption{MOND masses and implied M/L values for the total sample 
\label{t3}}
\begin{tabular}{|c|c|c|c|c|c|c|c|}
\hline
\ Galaxy & ${\rm M_d}$ & ${\rm (M/L)_d}$ & ${\rm M_b}$ & ${\rm (M/L)_b}$
  & ${\rm {M^*}/{L_B}}$ & ${\rm {M_t}/{L_B}}$ & ${\rm {M_t}/{L_H}}$ \\
\  &${\rm 10^{10}M_\odot}$ & &${\rm 10^{10}M_\odot}$ & & & & \\ 
\ (1) & (2) & (3) & (4) & (5) & (6) & (7) & (8) \\
\hline
  UGC 2885 & 25.1 &   1.3 &  5.7 & 3.9 & 1.5 &  1.7 &   \\
  NGC ${ \rm2841^*}$ & 24.1 &   3.5 & 6.9 & 4.3 & 3.7 &  4.0 &  1.9  \\
  NGC 5533 &  2.0 &   0.6 & 17.0 & 7.2 & 3.4 &  3.9 &  \\
  NGC 6674 &  2.5 &   1.8 & 15.5 & 2.9 & 2.6 &  3.2 &  \\
  NGC ${\rm 7331^*}$ &  8.6 &   3.8 &  4.7 & 1.5 & 2.5 &  2.7 &  0.8  \\
  NGC 5907 &  7.2 &   1.6 &  2.5 & 6.8 & 3.9 &  4.3 &  2.2  \\
  NGC 2998 &  5.4 &   0.9 &  2.9 & 4.3 & 1.2 &  1.7 &   \\
  NGC  801 &  3.5 &   0.7 &  6.5 & 2.5 & 1.4 &  1.7 &  \\
  NGC 5371 &  6.7 &   1.4 &  4.8 & 1.8 & 1.6 &  1.7 &  \\
  NGC 5033 &  8.8 &   4.6 &      &     & 4.6 &  5.1 &  2.5  \\
  NGC ${\rm 2903^*}$ &  5.5 &   3.6 &      &     & 3.6 &  3.8 &  2.7  \\
  NGC 3521 &  6.2 &   3.1 &  0.3 & 0.7 & 2.7 &      &   \\
  NGC 2683 &  3.0 &   6.4 &  0.5 & 2.8 & 5.8 &      &  \\
  NGC ${\rm 3198^*}$ &  2.3 &   2.6 &      &     & 2.6 &  3.3 &  3.6  \\
  NGC 6946 &  2.7 &   0.5 &      &     & 0.5 &  1.0 &   \\
  NGC ${\rm 2403^*}$ &  1.1 &   1.4 &      &     & 1.4 &  2.0 &  1.6  \\
  UGC  128 &  0.57 &   1.1 &     &     &  1.1 &  2.8 &  \\
  NGC ${\rm 6503^*}$ &  0.83 &  1.7 &      &     & 1.7  & 2.2  & 2.3 \\
  NGC 1003 &  0.30 &  0.2 &      &     & 0.2  & 0.7  & 2.5  \\
  NGC  247 &  0.40 &  1.1 &      &     & 1.1  & 1.5  & 2.3  \\
  M     33 &  0.48 &  0.65 &     &     & 0.6  & 0.8  & 1.4  \\     
  NGC 7793 &  0.41 &  1.20 &     &     & 1.2  & 1.5  & 2.8   \\
  NGC  300 &  0.22 &  0.73 &     &     & 0.7  & 1.2  &  \\
  NGC 5585 &  0.12 &  0.50 &     &     & 0.5  & 1.5  & 2.6  \\
  NGC 2915 &  0.25 &  6.9  &     &     & 6.9  & 9.7  &  \\
  UGC ${\rm 2259^*}$ &  0.22 &  2.1  &     &     & 2.1  & 2.6  &  \\
  NGC   55 &  0.10 &  0.23 &     &     & 0.2  & 0.5  &  \\
  NGC ${\rm 1560^*}$ &  0.034&  1.0  &     &     & 1.0  & 3.8  & 2.1  \\
  IC  2574 &  0.01 &  0.13 &     &     & 0.1  & 1.0  & 3.5  \\
  DDO  ${\rm 170^*}$ &  0.024 &  1.5 &     &     &  1.5 &  5.3 &  \\
  NGC ${\rm 3109^*}$ &  0.005 & 0.1  &     &     & 0.1  & 1.4  &  \\
  DDO  168 &  0.005 & 0.23 &     &     & 0.2  & 1.7  &  \\
  DDO  ${\rm 154^*}$ &  0.004 & 0.11 &     &     & 0.1  & 9.1  &  \\
\hline
\end{tabular}
\tablecomments{The asterisk denotes galaxies from the sample of BBS}
\end{flushleft}
\end{table}

\clearpage

\figcaption[ ]{MOND fits to the rotation curves of the sample galaxies.
The radius (horizontal axes) is given in kpc in all cases and the rotation 
velocity in km/s.
The points with error bars are the observations and the solid line is
the rotation curve determined from the distribution of light and neutral
hydrogen with the MOND formula.  The other curves are the Newtonian
rotation curves of the various separate components:  the long dashed curve 
is the rotation velocity resulting from a central bulge, if present;  the
short dashed line is the rotation curve of the gaseous disk (HI plus He);
the dotted curve is that of the luminous disk.  The free parameter(s) of
the fitted curve are the disk mass and, if present, the bulge mass.
The sample galaxies are shown in order of decreasing asymptotic
circular velocity.  \label{fig1}}

\figcaption[ ]{A log-log plot of ${\rm M^*/L_B}$ vs. the observed
asymptotic rotation 
velocity for the sample galaxies.  Here ${\rm M^*}$ is the total mass
of the stellar component (disk plus bulge) determined from the MOND fit.
\label{fig2}} 

\figcaption[ ]{Log of ${\rm M^*/L_B}$ of sample galaxies vs. the 
reddening corrected B-V color from the Third Reference Catalogue (de
Vaucouleurs et al. 1991).  Also shown (dashed line) are theoretical
${\rm M/L_B}$ from population synthesis models of Larson and Tinsley (1978). 
\label{fig3}}

\figcaption[ ]{A log-log plot of the B-band luminosity 
of the sample galaxies in units of $10^{10}$ ${\rm L_\odot}$
vs. the observed asymptotic rotational velocity (the B-band Tully-Fisher
relation). \label{fig4}}

\figcaption[ ]{A log-log plot of the H-band luminosity 
($10^{10}$ ${\rm L_\odot}$) vs. the observed 
asymptotic rotation velocity (the H-band Tully-Fisher relation).  Only
15 galaxies from this sample have measured H-band magnitudes. \label{fig5}}

\figcaption[ ]{A log-log plot of ${\rm M_t/L_B}$ vs. the observed
asymptotic rotation velocity.  Here ${\rm M_t = M^* + M_g}$. 
This is the total mass of 
the galaxy-- the mass of the stellar component determined from the MOND fit
plus ${\rm M_g}$, the mass of the gaseous component.  \label{fig6}}

\figcaption[ ]{A log-log plot of ${\rm M_t/L_H}$ vs. observed asymptotic
rotation velocity for those 15 sample galaxies with measured H-band
magnitudes.  The scatter in this measured M/L is comparable to the scatter
in the observed H-band Tully-Fisher relation.  \label{fig7}}

\figcaption[ ]{a) The MOND fit to the rotation curve of NGC 2841 assuming
that this galaxy is at its Hubble law distance ($H_o = 75$ km/s Mpc) of
9.5 Mpc.  The Newtonian rotation curves of the various components is shown
as in Fig.\ 1.  This is an example of an unacceptable MOND fit.  
The fitted bulge and disk mass-to-light ratios in the blue are 0.6 and
13 respectively.
b) The MOND fit to the rotation curve of NGC 2841 allowing
distance to be a free parameter.  The implied distance, 19.3 Mpc, is twice
the Hubble law distance.  Here the bulge and disk M/L values are
4.3 and 3.5 respectivley.  \label{fig8}}
 
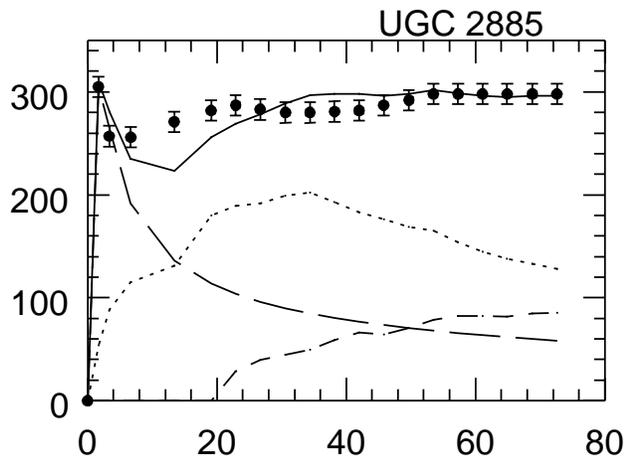
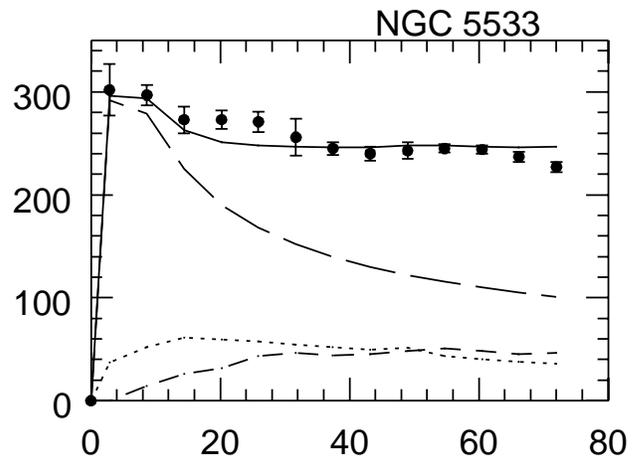
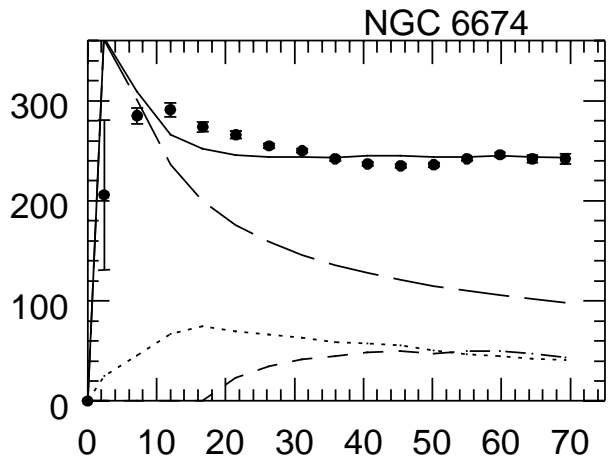
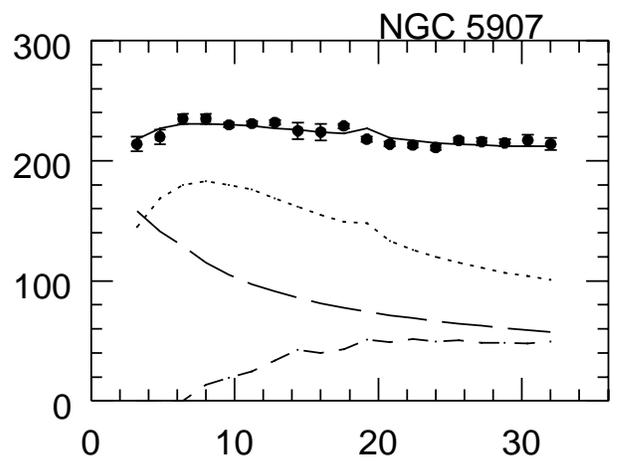
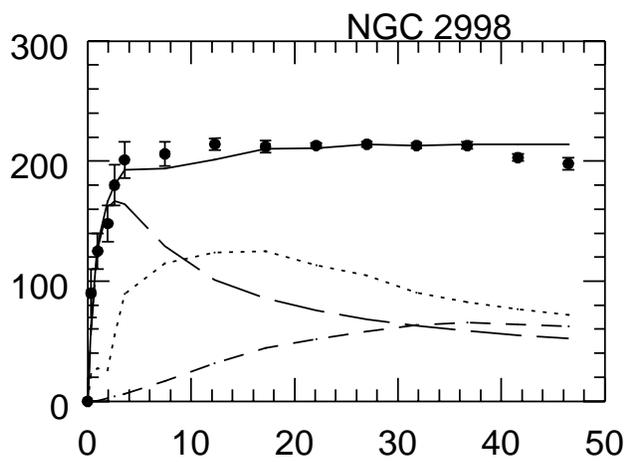
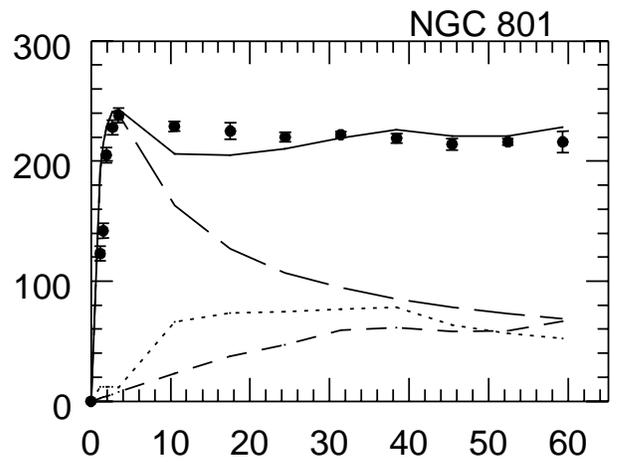
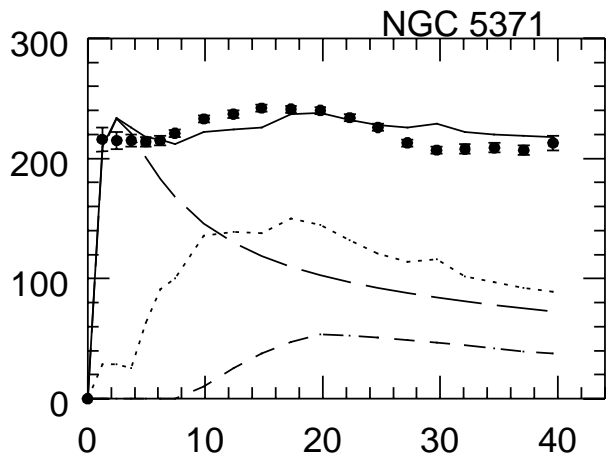
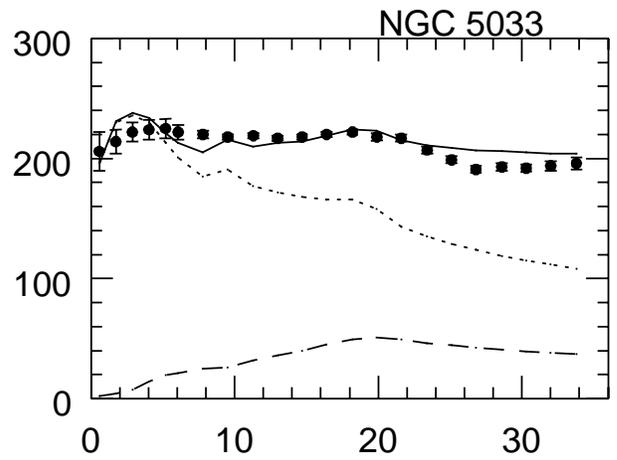
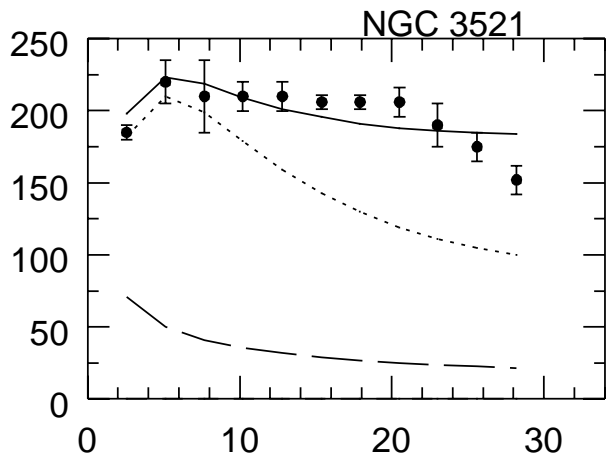
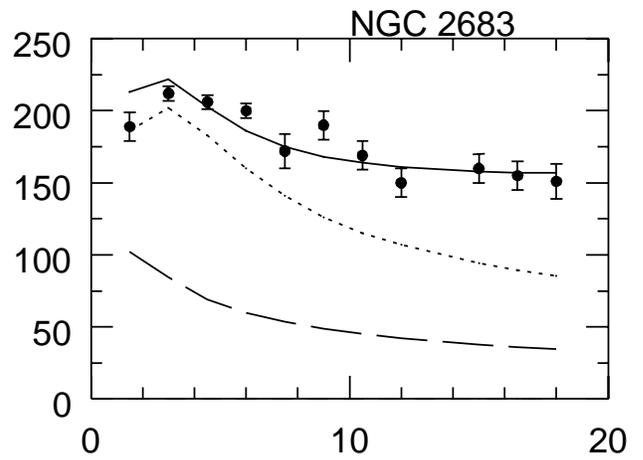
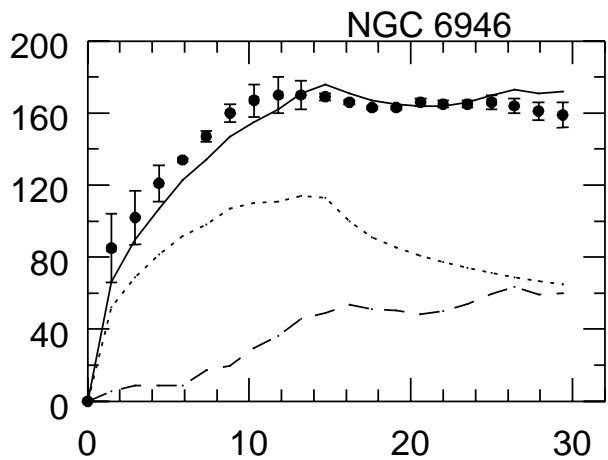
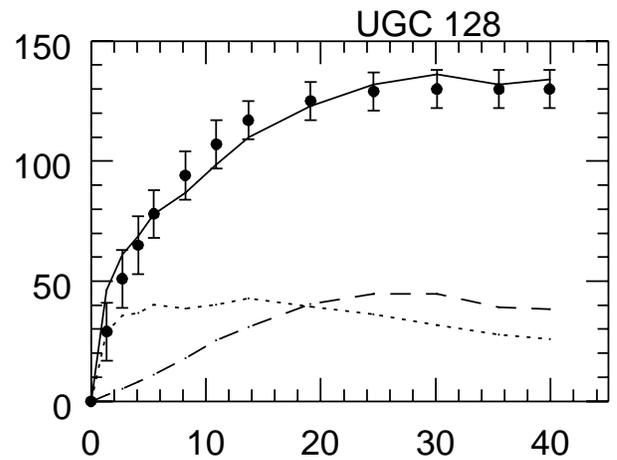
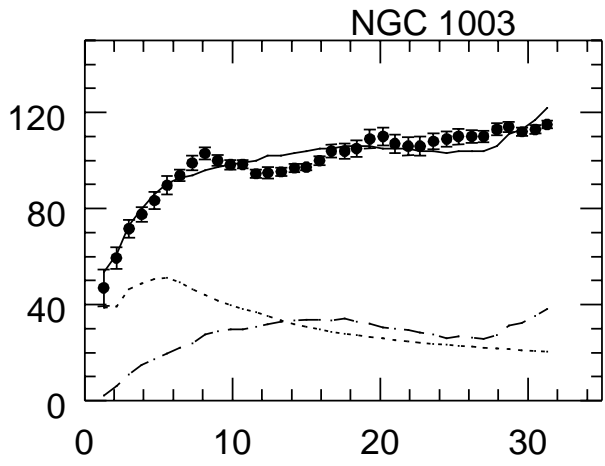
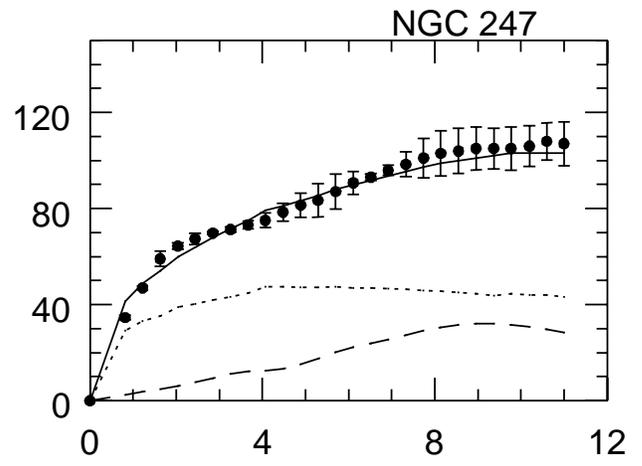
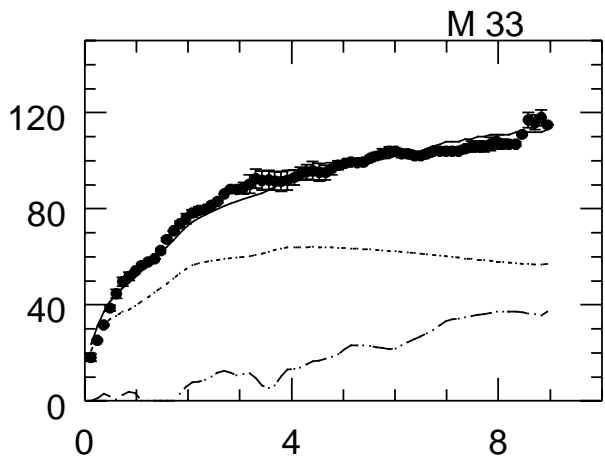
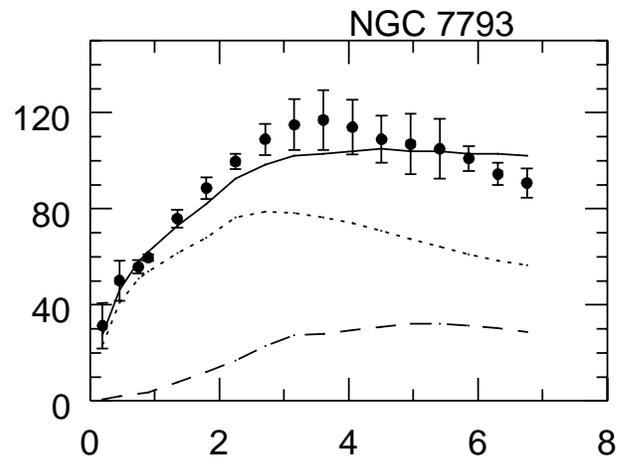
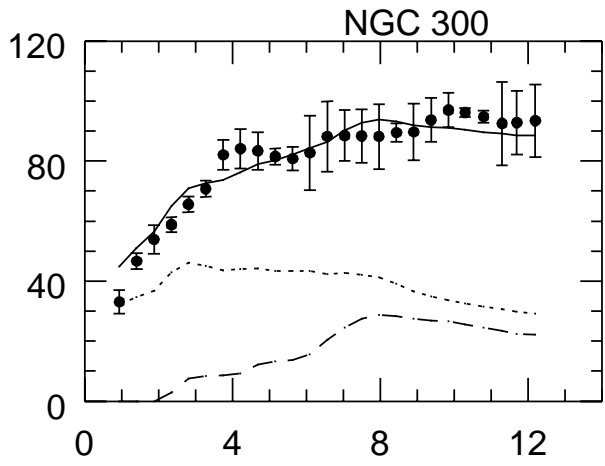
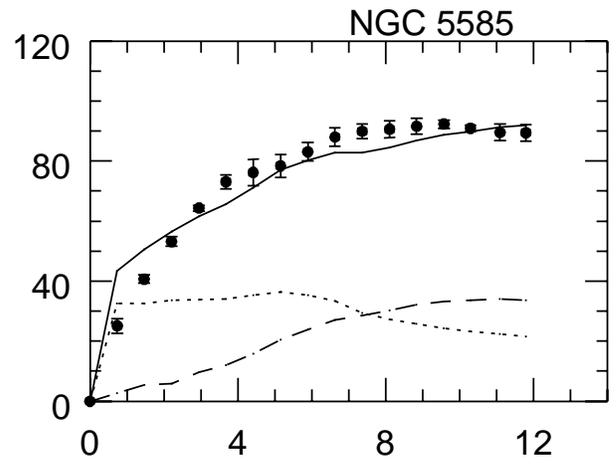
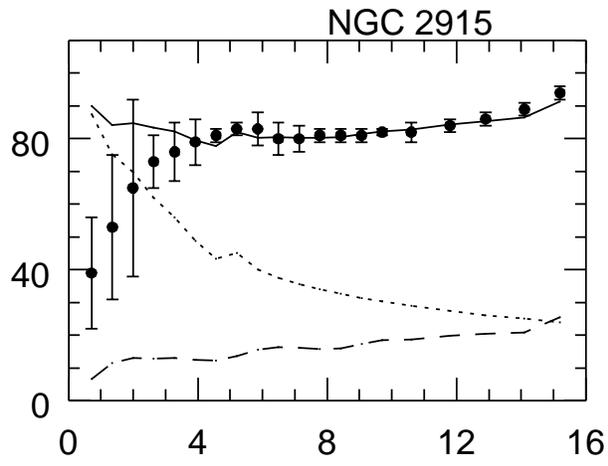
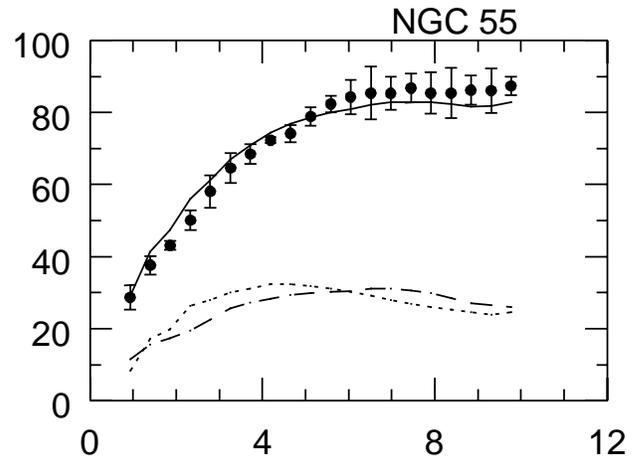
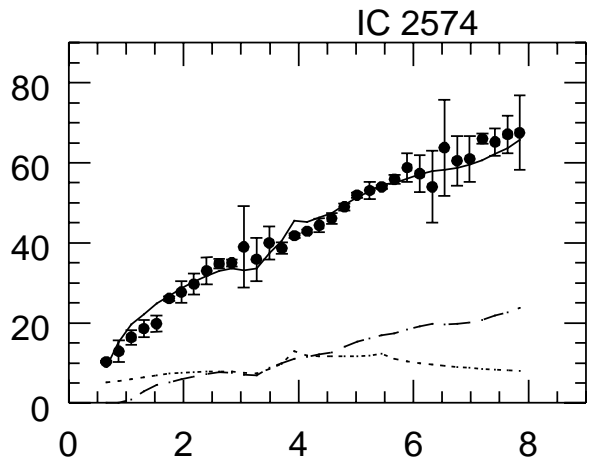
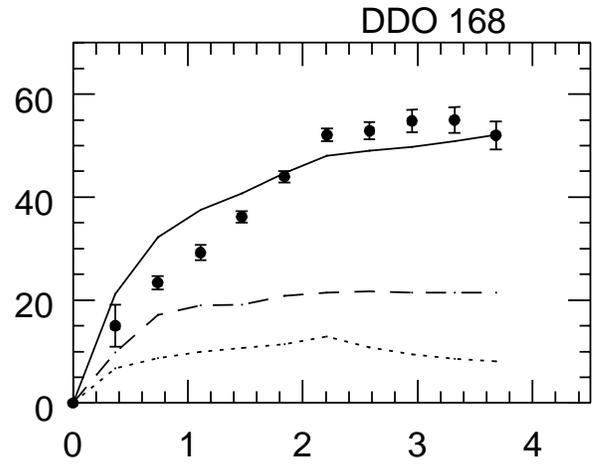
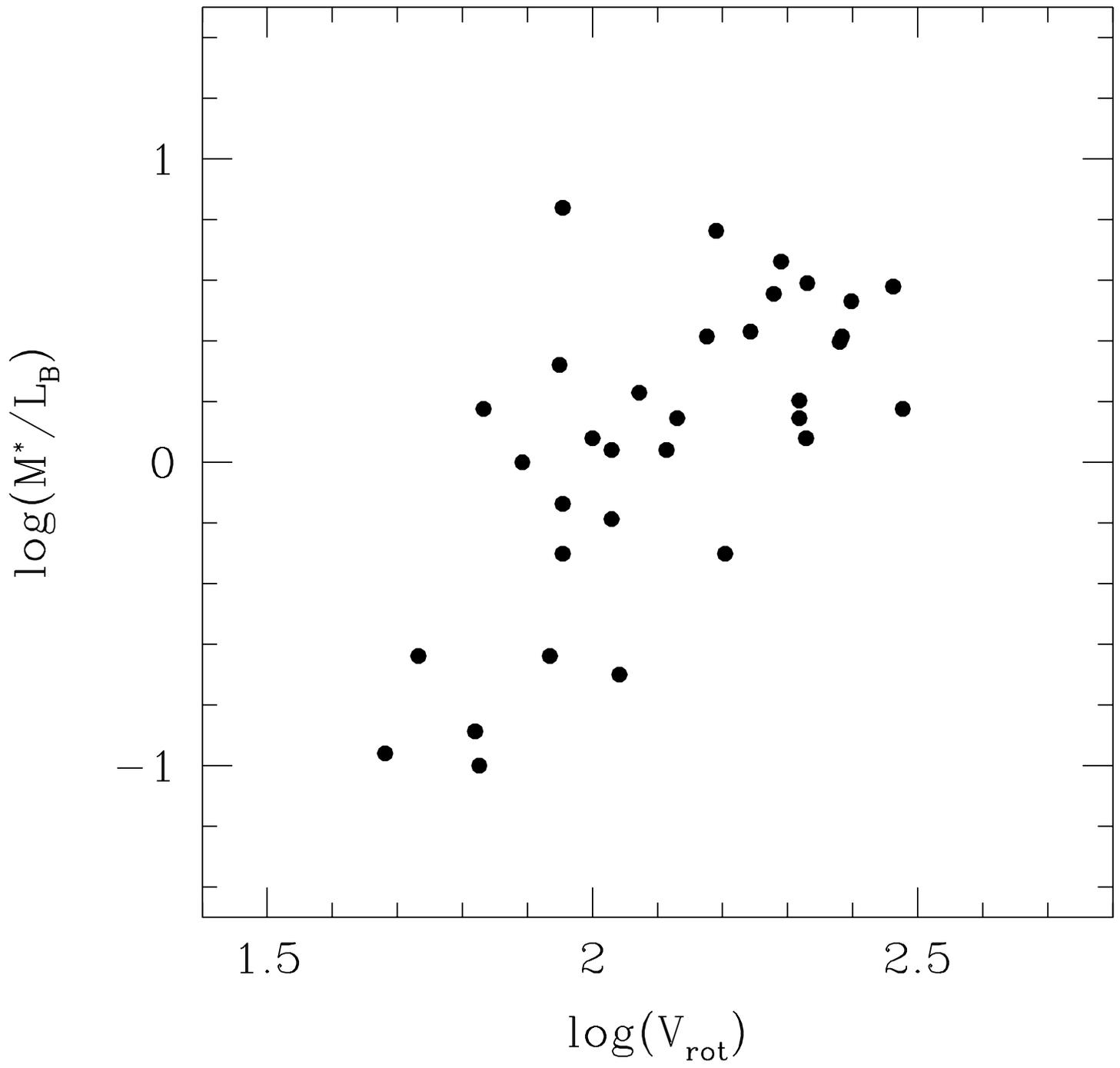
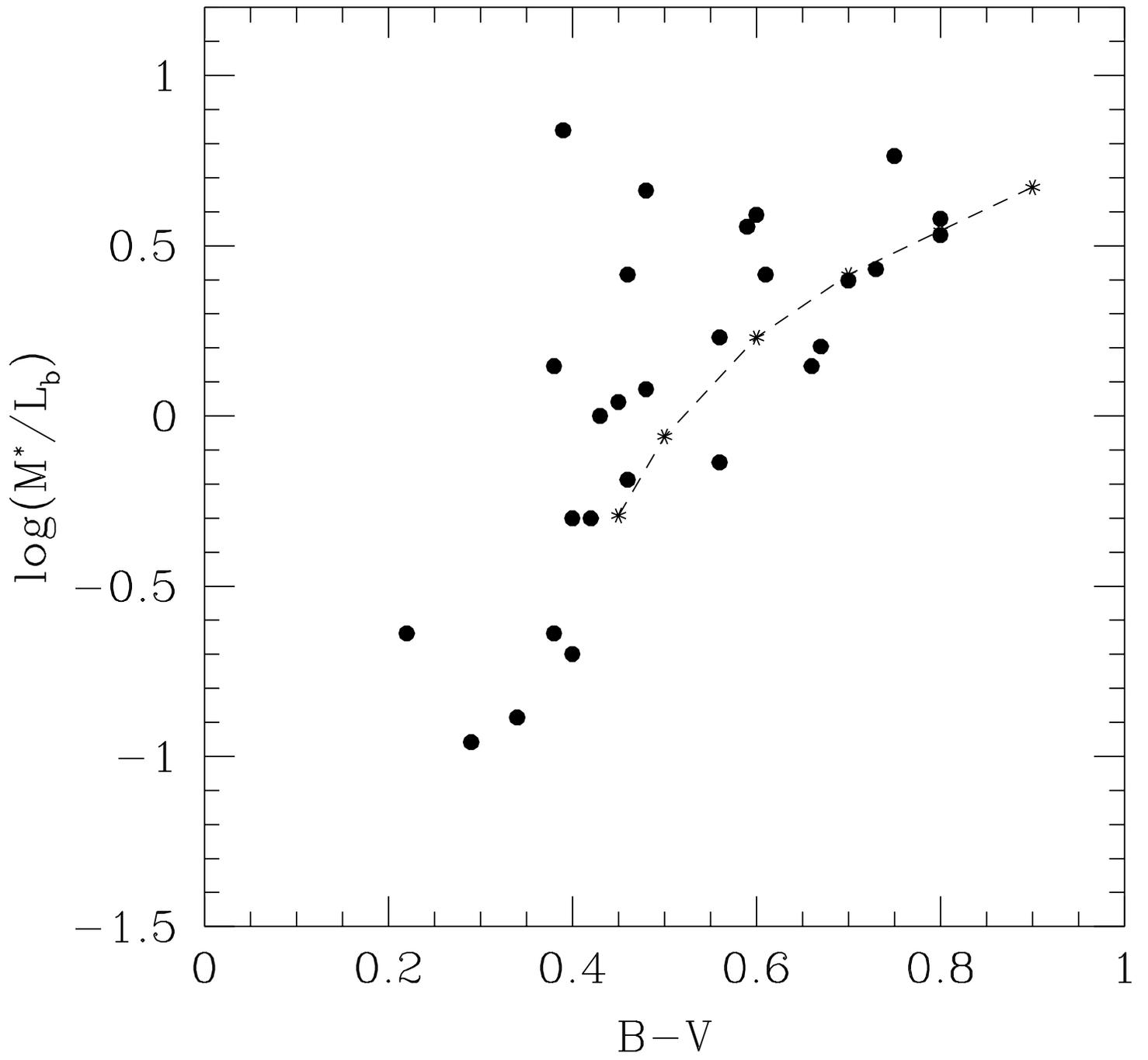
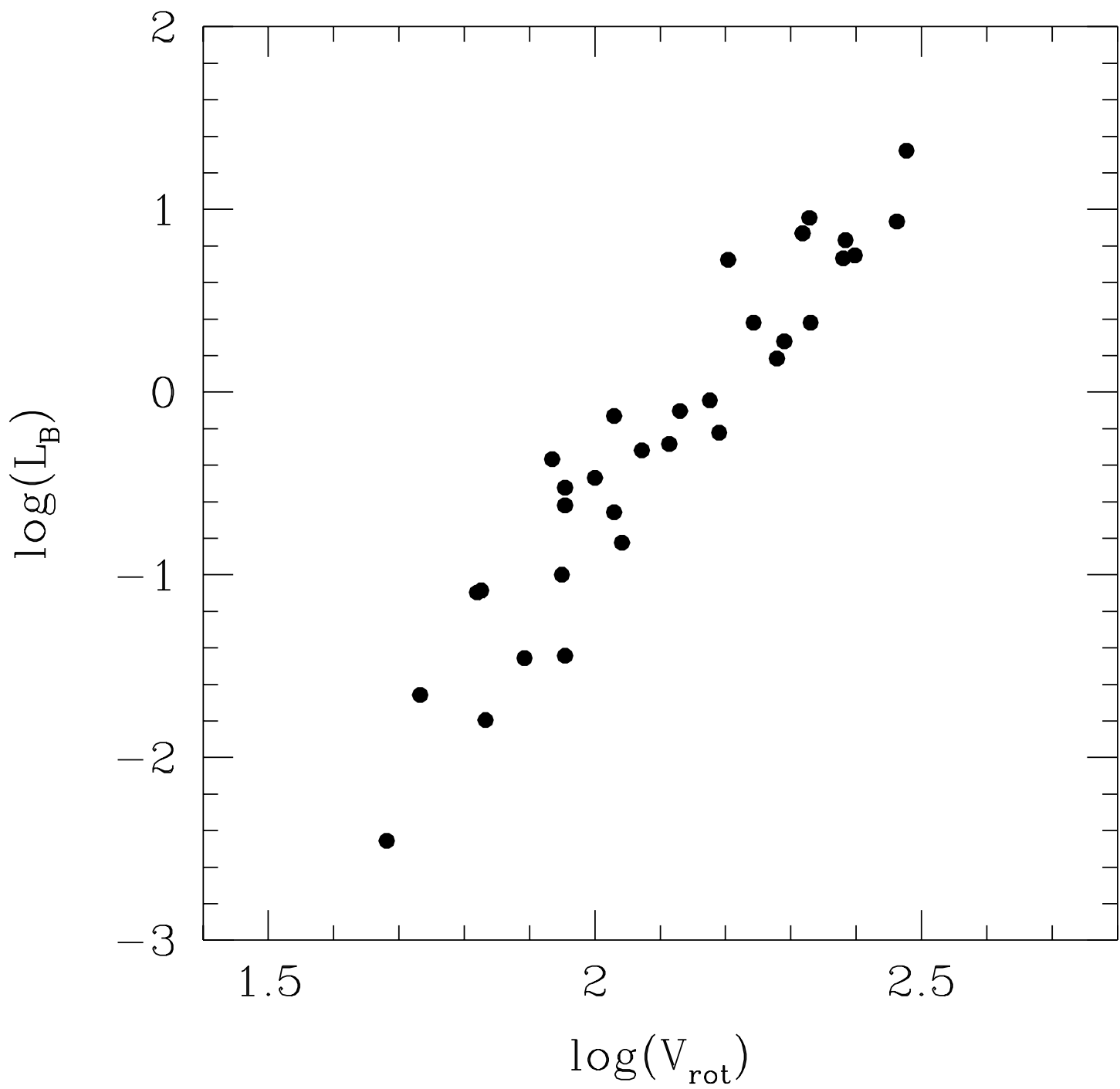
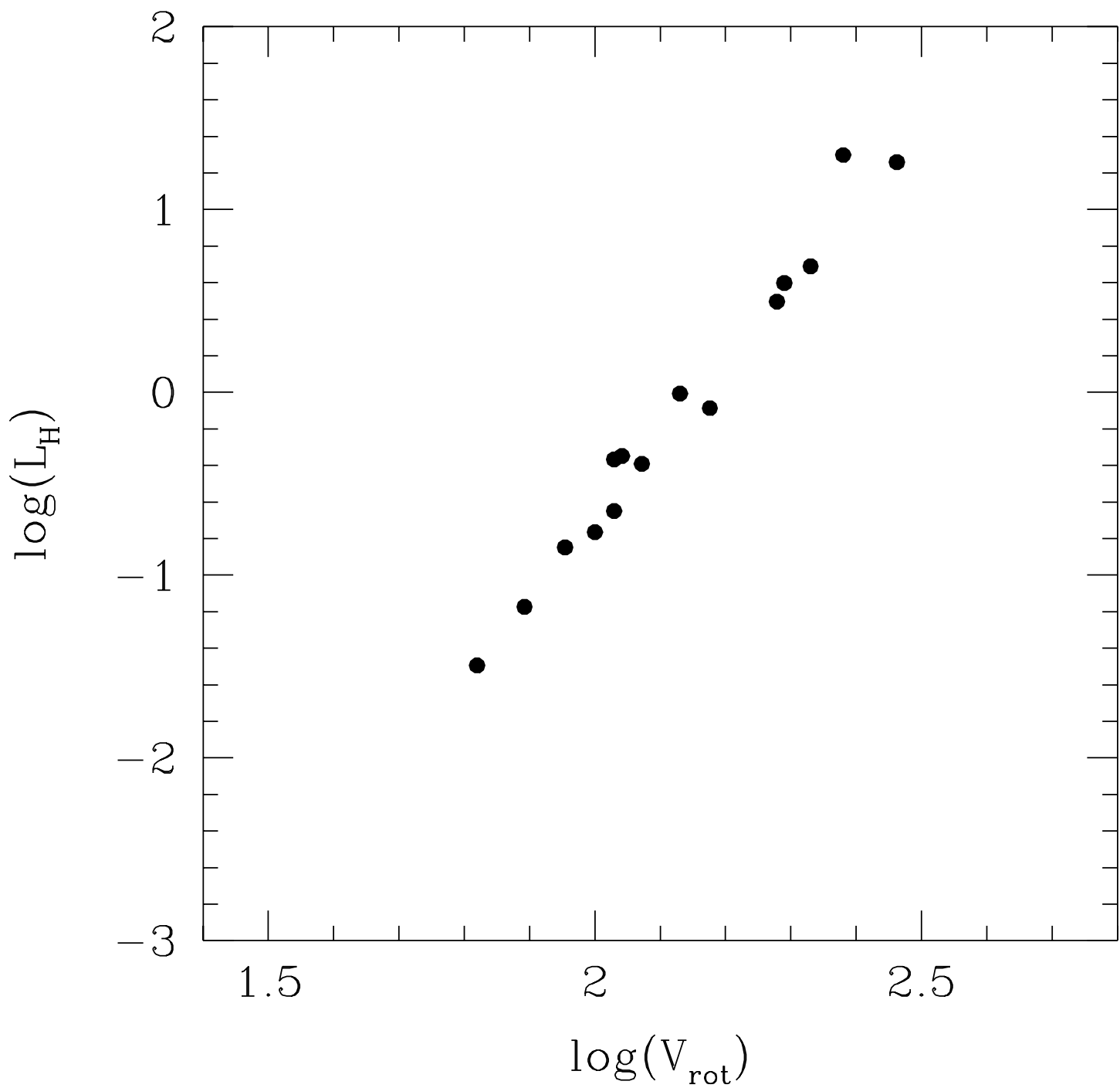
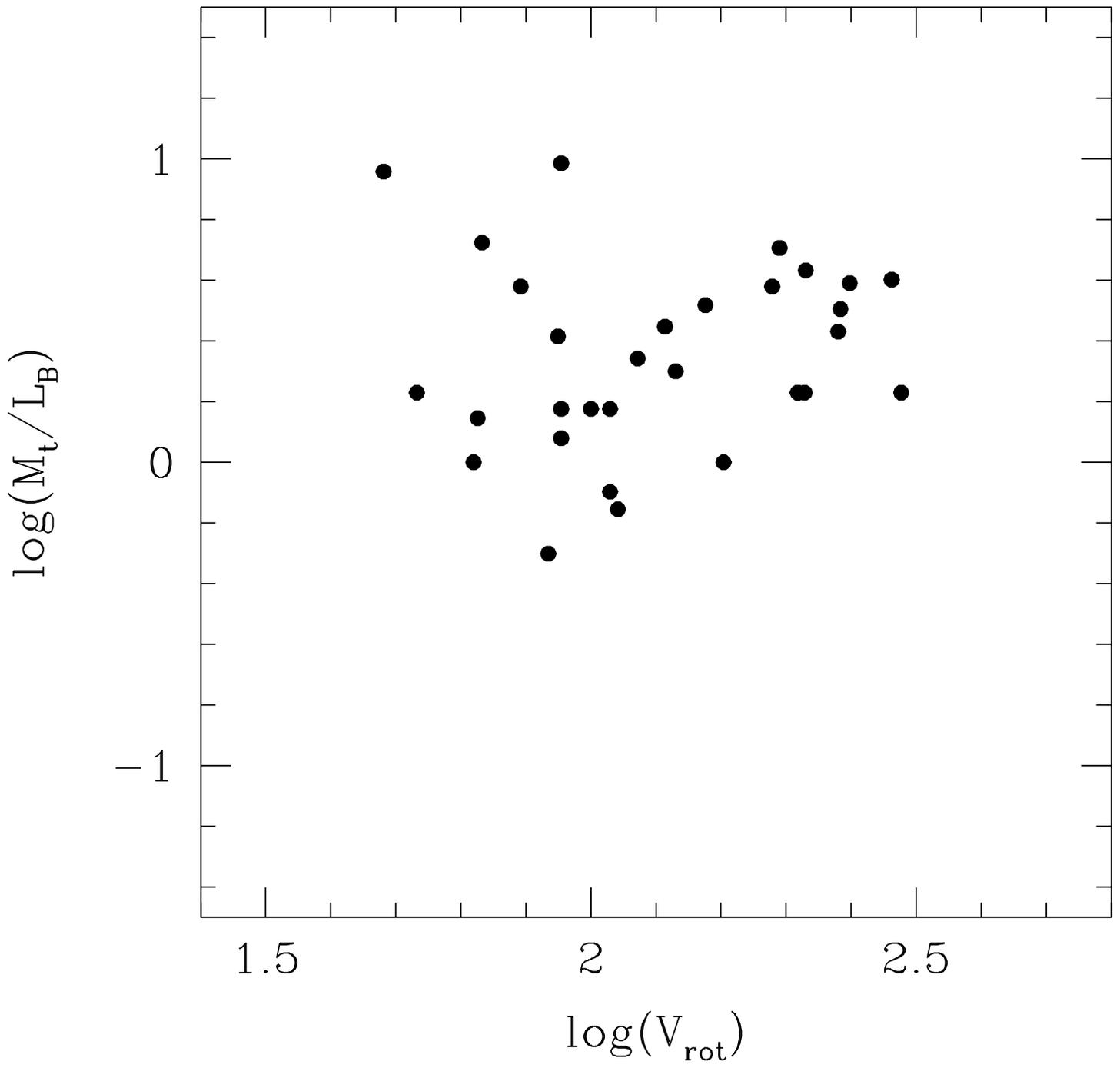
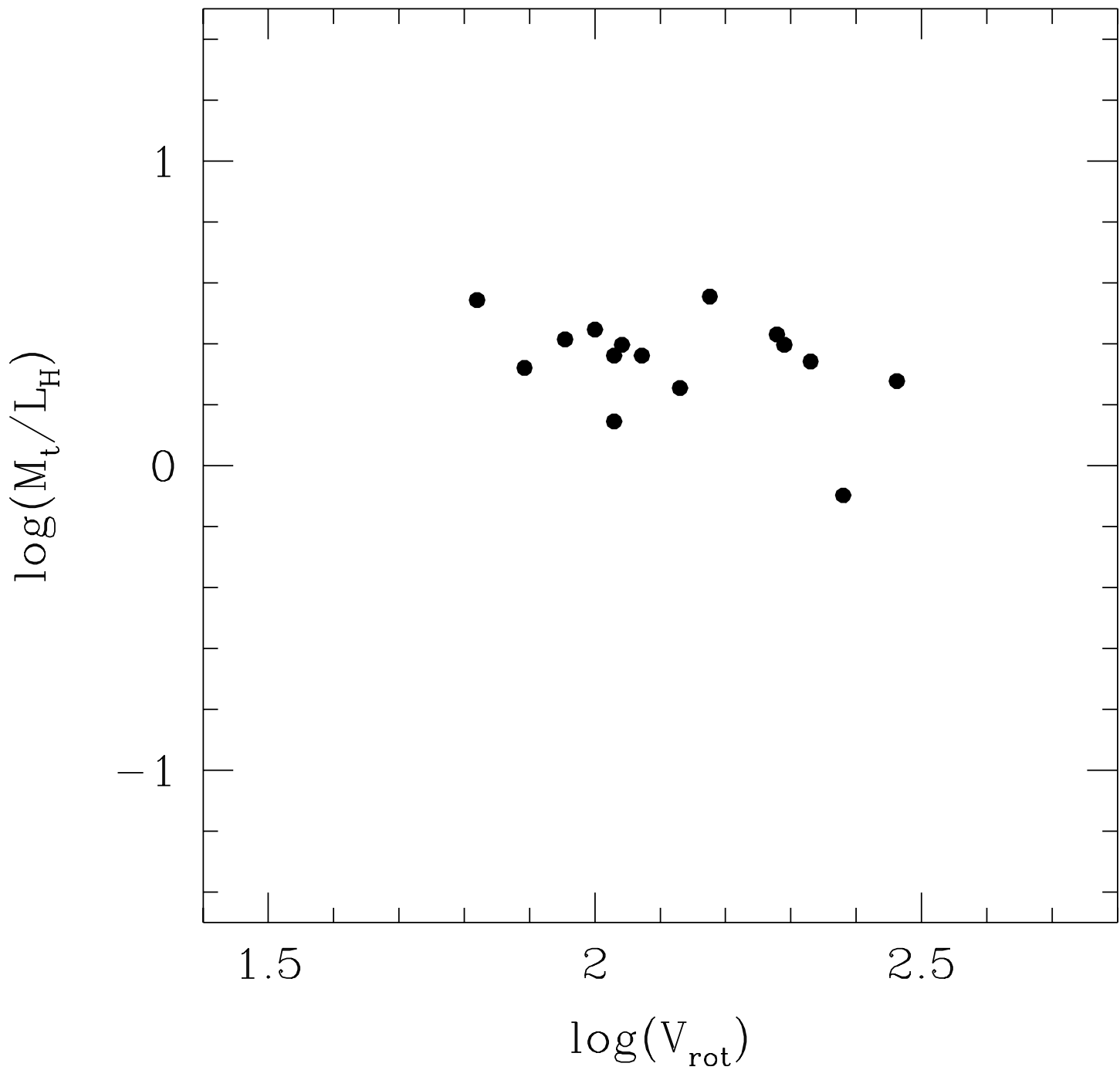
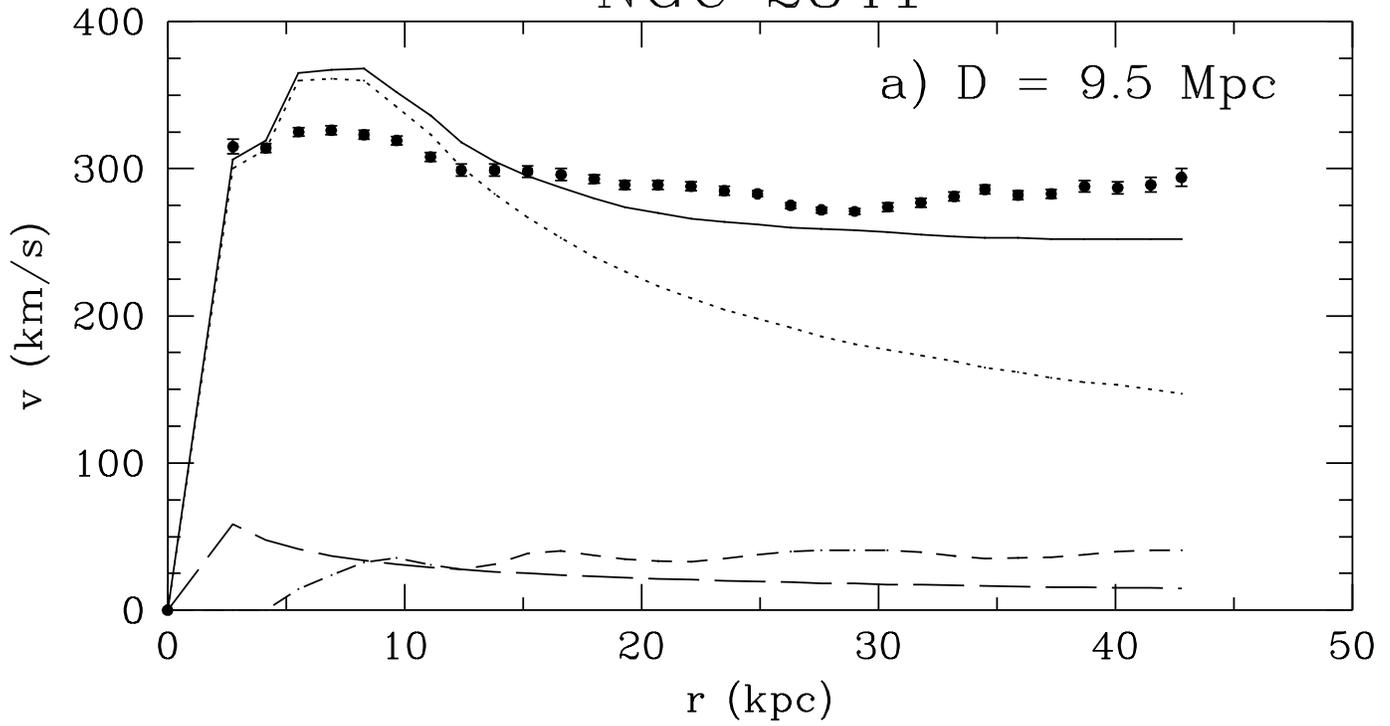
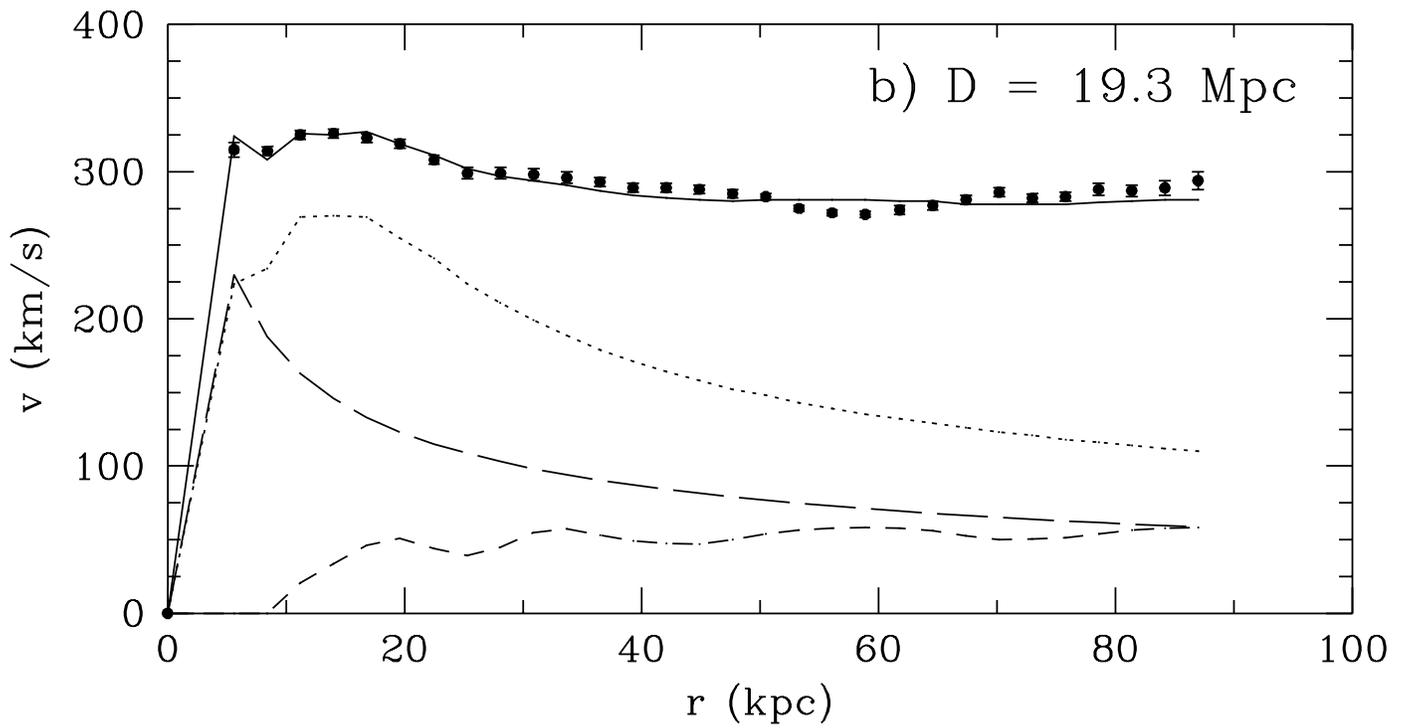
\end{document}